\documentclass[twocolumn,prb,aps,amsmath,amssymb,superscriptaddress,showpacs]{revtex4}
\usepackage{graphicx}
\usepackage{bm}
\usepackage[usenames,dvipsnames]{color}

\renewcommand\Im{\operatorname{Im}}

\newcommand\e{\mathrm{e}}

\newcommand\sign{\mathrm{sign}}

\newcommand\calE{\mathcal{E}}

\newcommand\calO{\mathcal{O}}

\newcommand\bfa{\mathbf{a}}
\newcommand\bfk{\mathbf{k}}

\newcommand\bfA{\mathbf{A}}
\newcommand\bfB{\mathbf{B}}
\newcommand\bfb{\mathbf{b}}
\newcommand\bfE{\mathbf{E}}
\newcommand\bfe{\mathbf{e}}
\newcommand\bfj{\mathbf{j}}

\newcommand\bfL{\mathbf{L}}

\newcommand\bfp{\mathbf{p}}
\newcommand\bfr{\mathbf{r}}
\newcommand\bfs{\mathbf{s}}
\newcommand\bfR{\mathbf{R}}
\newcommand\bfS{\mathbf{S}}

\newcommand{\bfalpha}{{\boldsymbol{\alpha}}}
\newcommand{\bfLambda}{{\boldsymbol{\Lambda}}}
\newcommand{\bfepsilon}{{\boldsymbol{\epsilon}}}

\newcommand{\bfSigma}{{\boldsymbol{\Sigma}}}
\newcommand{\bfsigma}{{\boldsymbol{\sigma}}}
\newcommand{\bfpi}{{\boldsymbol{\pi}}}

\newcommand{\HB}{H^{\mathrm{B}}}
\newcommand{\HFW}{H^{\mathrm{FW}}}

\usepackage{cancel}

\begin{document}
\title{Gauge invariance and relativistic effects in
photon absorption and scattering by matter}

\author{Nadejda Bouldi}
 \affiliation{Sorbonne Universit\'es,
   UPMC Univ Paris 06, 
   UMR CNRS 7590, 
  Mus\'eum National d'Histoire Naturelle, 
 \\ IRD UMR 206,
  Institut de Min\'eralogie, de Physique des Mat\'eriaux et de
 Cosmochimie, \\ 
 4 place Jussieu, F-75005 Paris, France}
 \affiliation{Synchrotron SOLEIL, 
  L'Orme des Merisiers, Saint-Aubin, BP 48,
  91192 Gif-sur-Yvette Cedex, France}
\author{Christian Brouder}
 \affiliation{Sorbonne Universit\'es,
   UPMC Univ Paris 06, 
   UMR CNRS 7590, 
  Mus\'eum National d'Histoire Naturelle, 
\\ IRD UMR 206,
  Institut de Min\'eralogie, de Physique des Mat\'eriaux et de
 Cosmochimie, \\ 
 4 place Jussieu, F-75005 Paris, France}
\date{\today}
\begin{abstract}
There is an incompatibility between gauge invariance and 
the semi-classical time-dependent perturbation theory commonly used to
calculate light absorption and scattering cross-sections.
There is an additional incompatibility between perturbation
theory and the description of the electron dynamics by a
semi-relativistic Hamiltonian.

In this paper, the gauge-dependence problem of
exact perturbation theory is described,
the proposed solutions are reviewed and it is concluded 
that none of them seems fully satisfactory.
The problem is finally solved by using the
fully relativistic absorption and scattering cross-sections
given by quantum electrodynamics.
Then, a new many-body Foldy-Wouthuysen transformation
is presented to obtain correct semi-relativistic 
transition operators. 
This transformation considerably simplifies the
calculation of relativistic corrections.
In the process, a new light-matter interaction term
emerges, called the spin-position interaction,
that contributes significantly to 
the magnetic x-ray circular dichroism of
transition metals.

We compare our result with the ones obtained 
by using several semi-relativistic time-dependent
Hamiltonians.
In the case of absorption, the final formula  agrees 
with the result obtained from one of them.
However, the correct scattering 
cross-section is not given by any of the semi-relativistic
Hamiltonians.
\end{abstract}
\maketitle

\section{Introduction}

There is a well-known conflict between time-dependent
perturbation theory and gauge invariance
(see section~\ref{gauge-inv-sect} for a 
non exhaustive list of references). 
Indeed, $|\langle \phi_n|\psi(t)\rangle|^2$ gives
the probability to find the system
described by the state $|\psi(t)\rangle$
in the eigenstate $|\phi_n\rangle$ at time $t$,
where $|\psi(t)\rangle$ is a solution of
the time-dependent Schr\"odinger equation
for the Hamiltonian
$H(t)=H_0+H_1(t)$, while $|\phi_n\rangle$ is
an eigenstate of the time-independent Hamiltonian
$H_0$. A time-dependent gauge transformation
of $H(t)$ will be assigned to
$H_1(t)$ but not to $H_0$, which
must remain independent of time. 
The transition probability is then modified because
the state $|\psi(t)\rangle$ is gauge-transformed
and not the state $|\phi_n\rangle$
(see section~\ref{gauge-inv-sect} for a 
more detailed argument).

As a (not so well-known) consequence,
the absorption and scattering cross-sections derived
by semi-classical arguments are not gauge
invariant. Since a basic principle of quantum physics  states
that an observable has to be gauge invariant to be
physically meaningful, we meet a serious difficulty:
``Until this problem is understood,
therefore, it seems that no calculation can be
trusted at all.''\cite{Kobe-78}

There is a further conflict between time-dependent 
perturbation theory and semi-relativistic physics.\cite{Yang-82-3}
The semi-relativistic approximation
of $|\phi_n\rangle$ is obtained by applying
to it the time-independent Foldy-Wouthuysen
transformation: $|\phi_n^{\mathrm{FW}}\rangle=U_{H_0}
|\phi_n\rangle$. The semi-relativistic
approximation of $|\psi(t)\rangle$ is derived
from the time-dependent Foldy-Woutuysen
transformation $|\psi^{\mathrm{FW}}(t)\rangle=U_{H(t)}
|\psi(t)\rangle$. 
Since $U_{H(t)}\not=U_{H_0}$, the 
transformed transition probability
$|\langle\phi_n^{\mathrm{FW}}|\psi^{\mathrm{FW}}(t)\rangle|^2$
is not equal to 
$|\langle \phi_n|\psi(t)\rangle|^2$, even if the
Foldy-Wouthuysen transformations $U_{H_0}$
and $U_{H(t)}$ are known to all orders.

In this paper we discuss and solve these two conflicts.
In a nutshell, the gauge problem is solved by
deriving relativistic absorption and scattering
cross-sections from quantum electrodynamics
instead of the usual semi-classical argument where the 
incident light wave is described by a time-dependent
potential.
The semi-relativistic problem is solved by 
applying a new many-body Foldy-Wouthuysen transformation
to the relativistic cross-sections instead
of describing the dynamics of the system
with a semi-relativistic Hamiltonian.
The final result is a semi-relativistic
absorption and scattering cross-section
involving a new term that couples the spin
and the position operators. 
In a companion paper, we show that this
new term contributes significantly
to the x-ray absorption of magnetic materials.\cite{Bouldi-XMCD}

We now describe the outline of this paper.
Section 2 discusses the gauge transformation
of transition probabilities and reviews the 
solutions to the gauge-dependence problem
proposed in the literature. Since none of them
was widely accepted, we turn to the quantum
electrodynamics framework in section 3,
where we derive the relativistic 
electric dipole, quadrupole and 
magnetic dipole relativistic transition operators.
In section 4, we derive a new many-body 
Foldy-Wouthuysen transformation that we
apply to the transition operators.
They are used to obtain semi-relativistic
absorption and scattering cross-sections
in sections 5 and 6, where a new spin-position
term is derived.
In section 7, the conflict between time-dependent
perturbation theory and semi-relativistic 
methods is described and illustrated by
the calculation of spin-position term using
four different semi-relativistic Hamiltonians
commonly used in the literature.
The conclusion presents possible extensions
of the present work.

\section{Gauge invariance}
\label{gauge-inv-sect}
The gauge invariance of the
absorption and scattering cross-sections of light
is a long-standing problem.
It started in 1952 when Willis Lamb calculated
the spectrum of Hydrogen in two gauges and obtained 
different results.\cite{Lamb-52} This
gave rise to a long series of papers
up to this day.~\cite{Grant-74,Grant-75,Yang-76,Zeyher-76,Forney-77,Kobe-78,%
Aharonov-79,Epstein-79,Grynberg-79,Haller-79,Olariu-79,%
Kobe-80,Kobe-80-2,Leubner-80,%
Aharonov-81,Shirokov-81,%
Kobe-82,Kobe-82-2,Aharonov-82,Yang-82-2,Yang-82-3,%
Yang-82,Feuchtwang-82,Kazes-82,%
Aharonov-83,Lee-83,Yang-83,%
Au-84,Feuchtwang-84,Feuchtwang-84-2,Kobe-85,Adu-Gyamfi-86,Yang-86,%
Kobe-87,Lamb-87,Arrighini-88,Durante-88,Yang-88,%
Haller-94,%
Woolley-98,Woolley-99,Woolley-00,Savukov-00,Stewart-00,%
Qian-02,Woolley-02,Rzazewski-04,Qian-08,%
Stokes-13,Bandrauk-13,Mandal-16}
In 1987, the same Lamb (then Nobel prize winner)
still considered this as ``one of the
outstanding problems of modern quantum optics.''~\cite{Lamb-87}

We quickly describe the meaning of gauge invariance
and then consider its failure in semi-classical perturbation theory.

\subsection{The principle of gauge invariance}
The two homogeneous Maxwell equations
$\nabla\times \bfE + \dot\bfB = 0$ and
$\nabla\cdot\bfB = 0$, where the dot denotes time derivative,
imply the local existence of a
vector potential $\bfA$ and a scalar potential $\Phi$ such
that $\bfB=\nabla\times\bfA$ and
$\bfE=-\nabla \Phi - \dot\bfA$.
We denote $A=(\Phi,\bfA)$.
The same $\bfE$ and $\bfB$ are obtained from the
potentials $A'=(\Phi-\dot\Lambda,\bfA+\nabla\Lambda)$,
that we also denote $A'=A-\partial\Lambda$, where $\Lambda$
is any smooth function of space and time.
In classical electromagnetism, gauge invariance means that 
the physics described by $A$ and $A'$ is the same.

In quantum mechanics, consider a non-relativistic 
Hamiltonian
\begin{eqnarray*}
H_A &=& 
\frac{(\mathbf{p} - e \mathbf{A})^2}{2m}
   + e \Phi,
\end{eqnarray*}
or a relativistic (Dirac) Hamiltonian
\begin{eqnarray*}
H_A &=& c {\boldsymbol{\alpha}}\cdot(\mathbf{p}-e\mathbf{A})
+ mc^2\beta  + e\Phi,
\end{eqnarray*}
where $\bfalpha=(\alpha_x,\alpha_y,\alpha_z)$ and
$\beta$ are the Dirac matrices.
Both Hamiltonians are of the form
$H_A=f(\bfp-e\bfA)+e\Phi$, where $f$ is some function.
For such Hamiltonians it can be checked that
$M_\Lambda^\dagger(i\hbar\partial_t-H_{A'})M_\Lambda=i\hbar\partial_t-H_A$,
where $M_\Lambda=e^{ie\Lambda/\hbar}$.
As a consequence, if $\psi$ is a solution of the time-dependent
Schr\"odinger equation
$i\hbar\dot\psi=H_A\psi$, then
$\psi'=M_\Lambda \psi$ is a solution of
$i\hbar\dot\psi'=H_{A'}\psi'$.

In quantum mechanics, a gauge transformation consists in both a change
of the potentials and a change in the phase of the wavefunctions.
An observable $O_A$ depending on the electromagnetic
potential $A$ is said to be \emph{gauge invariant}
if $M_\Lambda^\dagger O_{A'} M_\Lambda=O_A$ for every function
$\Lambda(t,\bfr)$. An observable must be gauge invariant
to
be considered a \emph{true physical quantity}.\cite{Cohen-QED}

The principle of gauge invariance 
 has become a cornerstone of particle physics.
Since general relativity can also be considered as
a gauge theory~\cite{Blagojevic}, it may be safely
said that gauge invariance was the guiding
principle of most of the fundamental physics of the 
twentieth century.
Therefore, we need to check that the
cross-section formulas are gauge invariant
to ensure their true physical nature.

Note that the time-dependent Dirac or Schr\"odinger
equations are always gauge invariant but the
time-independent ones are not because $H_A$ is not
gauge invariant due to the the
scalar potential $\Phi$. Indeed, under a gauge
transformation $\Phi$ becomes $\Phi-\dot{\Lambda}$
and the term $\dot{\Lambda}$ cannot be compensated
for in the absence of a time derivative.

\subsection{Gauge dependence of transition probabilities}
In time-dependent perturbation theory, a system is
assumed to be in the ground state $|\phi_g\rangle$ 
of a time-independent Hamiltonian $H_{a_0}$. Then, at
time $t_0$, an electromagnetic wave represented
by the time-dependent potential $a$ is added to
the system (with total potential
 $A=a_0+a$), which is represented at time $t$
by the state $|\psi(t)\rangle$.
A good way to take both the initial state and the dynamics
into account is to use the evolution operator
$U_{A}(t,t_0)$, which is the solution of
$i\hbar\partial_t U_{A}(t,t_0)=H_{A}(t)U_{A}(t,t_0)$
with the boundary condition $U_A(t_0,t_0)=1$.
Thus, $|\psi(t)\rangle=U_{A}(t,t_0)|\phi_g\rangle$.
The probability of a transition to the eigenstate
$|\phi_n\rangle$ of $H_{a_0}$ at time $t$ is
\begin{eqnarray}
P_{ng}(t) &=& |\langle \phi_n|\psi(t)\rangle|^2
= |\langle \phi_n |U_{A}(t,t_0)|\phi_g\rangle|^2.
\label{Png}
\end{eqnarray}
Since we want to ensure that the breakdown of gauge invariance
is not due to an approximation, we work with exact
(i.e. to all orders) perturbation theory.
If we carry out a gauge transformation of the perturbation
from $a$ to $a'=a-\partial \Lambda$, then 
the evolution operator becomes~\cite{Kobe-85}
\begin{equation}
U_{A'}(t,t_0)=M_\Lambda(t)U_{A}(t,t_0)M_\Lambda^\dagger(t_0),
\end{equation}
where $A'=a_0+a'$.

Therefore 
$\langle\phi_n|\psi'(t)\rangle=\langle
\phi_n|U_{A'}(t,t_0)|\phi_g\rangle$ is
\begin{eqnarray*}
\int d\bfr d\bfr'\phi_n^*(\bfr) e^{ie\Lambda(\bfr,t)}
U_A(\bfr,t;\bfr',t_0) e^{-ie\Lambda(\bfr',t_0)}
  \phi_g(\bfr'),
\end{eqnarray*}
which is generally different from $\langle\phi_n|\psi(t)\rangle$
since $\Lambda(\bfr,t)$ is an arbitrary function
(take for example $\Lambda(\bfr,t)=\bfr\cdot\bfk(t)$, where
$\bfk(t)$ is an arbitrary function of time).
Therefore, in general, $P'_{ng}(t) \not=P_{ng}(t)$
and the transition probabilities calculated in the two gauges
are different.
Moreover, since the transition rate entering cross-sections
is the derivative of the transition probability with
respect to time,~\cite{Bohm} 
the arbitrariness of the transition
rate is increased by the fact that an
arbitrary function $\Lambda(\bfr,t)$ enters the integrand.
Indeed, several papers evaluate the 
discrepancy between 
the probability calculated with two different gauges,
and they find that it
is generally not small.\cite{Lamb-52,Durante-88,Stokes-13}
By properly choosing $\Lambda$, the discrepancy 
can even be made arbitrary large.\cite{Rzazewski-04}

The absence of gauge invariance is due to the
fact that the operator is transformed but
not the eigenstates of $a_0$. This is called a
\emph{hybrid transformation} in the
literature.\cite{Haller-79}

\subsection{Proposed solutions}
The lack of gauge invariance of transition probabilities
is an alarming problem to which several solutions have been proposed.
Since no clear consensus appears to have
emerged,\cite{Stewart-00} we present a
critical review of these solutions.

The first one, called the \emph{consistent procedure},
was proposed by Forney and coll.
and Epstein.\cite{Forney-77,Epstein-79}
It is based on the observation
that, if instead of gauge-transforming $a$ we transform the potential
$a_0$ of the 
initial Hamiltonian to get $H_{a_0'}$, where
$a_0'=a_0-\partial\Lambda$,
then the evolution operator becomes again
$U_{A'}(t,t_0)$ 
(because $a_0'+a=a_0+a-\partial\Lambda=A'$)
but the eigenstates $|\phi_g\rangle$ and $|\phi_n\rangle$
are also transformed into time-dependent states
$|\phi'_g\rangle=M_\Lambda(t_0) |\phi_g\rangle$ and
$|\phi'_n\rangle=M_\Lambda(t) |\phi_n\rangle$.
Therefore, the transition probability is now conserved.
In other words, gauge invariance is lost if we 
subtract $\partial\Lambda$ from the perturbation 
but not if we subtract it from the unperturbed
Hamiltonian $H_0=H_{a_0}$.

Since the reference states $|\phi_n\rangle$ become time-dependent,
we leave the standard framework of time-dependent
perturbation theory where the initial Hamiltonian
$H_0$ does not depend on time. 
Moreover, it is not physically clear why the
gauge transformation should be applied to $H_0$
and not to the perturbation. 

In spite of these difficulties, 
many authors proposed to use
the consistent procedure.
However, as noticed by Yang,\cite{Yang-82-3}
this does not really solve the problem because, 
if we start the calculation with the
initial potential $H_{a_0}$ and the perturbation
$a'$, the transition probability is
$P'_{ng}(t)$. If we then use the consistent procedure
to come back to the perturbation $a$, then we still
find $P'_{ng}(t)$ and we do not recover
the result $P_{ng}(t)$. In other words, the
transition probability is now gauge invariant 
(in the sense that a change of gauge does not
modify the result) but it
is gauge-dependent (in the sense that the result
depends on the gauge we use in the perturbation 
to start the calculation).
This gauge dependence would be a serious problem 
because we would have to select the ``true'' physical
gauge for the perturbation.


A second solution appeared in a series of papers starting in 
1976,\cite{Yang-76,Kobe-80,Kobe-80-2,Yang-82-2,Yang-82-3,%
Yang-82,Kobe-82-2,Yang-83,Kobe-85,Yang-86,Yang-88}, where
Yang and collaborators proposed to define a gauge
invariant transition probability.
His idea is to start from the gauge-invariant
(but time-dependent) initial Hamiltonian
\begin{eqnarray}
H_0(t) &=& \frac{(\bfp-e\bfa_0-e\bfa(t))^2}{2m}
   + \mathbf{e}V,\label{HAaV}
\end{eqnarray}
where $V$ describes the electron-electron and electron-nuclear
interactions so that $H=H_0+e\phi$:
the perturbation is only the scalar potential $\phi$.
Then, the Hamiltonian $H_0(t)$
is diagonalized at every time $t$:
$H_0(t)|\phi_n(t)\rangle=E_n(t)|\phi_n(t)\rangle$ and
the transitions are calculated between the
time-dependent
states $|\phi_n(t)\rangle$. The corresponding transition
probabilities are indeed gauge invariant.
This solution has been used
up to this day,\cite{Arrighini-88,Qian-08,Mandal-16} 
although it was also strongly 
criticized.\cite{Aharonov-79,Leubner-80,Aharonov-81,%
Feuchtwang-82,Kazes-82,Aharonov-83,Au-84,Feuchtwang-84,%
Feuchtwang-84-2,Feuchtwang-85,Feuchtwang-86}
The main arguments against Yang's interpretation are:
(i) the quantity $E_n(t)$ is not physical
because you cannot measure an energy at a given time with
arbitrary precision; (ii) the time-dependent states 
$|\phi_n(t)\rangle$
can be neither prepared nor detected; 
(iii) the term $V$ in Eq.~(\ref{HAaV}) should be
removed from $H_0(t)$ because it is a scalar potential
and, as such, not gauge invariant. But if $V$ is removed,
then $H_0(t)$ is so far from the true Hamiltonian
that perturbation theory is no longer valid.

Following Goldman,\cite{Goldman-77}
Feuchtwang, Kazes and coll. proposed the following
alternative solution.\cite{Feuchtwang-82,Kazes-82,Kazes-83,%
Feuchtwang-84,Feuchtwang-84-2}
They started from the well-known
fact that the equations of motion of a Lagrangian
are not modified by the addition of the total
time derivative of a function.~\cite{Cohen-QED}
Thus, two Lagrangians that differ by a total time
derivative are equivalent.\cite{WeinbergQFT}
Then, they remark that the addition of a
total time derivative $e\dot\Lambda$ to the Lagrangian
induces a gauge transformation $A\to A-\partial\Lambda$
of the Hamiltonian.\cite{Kazes-83,Cohen-QED,Qian-02}
Finally, they use such a total derivative to
compensate for the electric potential that is the
cause of the gauge variance of the Hamiltonian.
However, it is difficult to distinguish this
procedure from picking up a specific gauge,
namely the Weyl or temporal gauge where the
scalar potential vanishes. 
We can conclude this short review by stating that
no solution was found fully satisfactory.

To determine when gauge invariance can be achieved
at the first order of perturbation theory,
we consider a Dirac Hamiltonian in two gauges 
$A$ and $A'=A-\partial\Lambda$ and
we calculate the difference
\begin{eqnarray*}
\langle \psi| H_A- H_{A'}|\psi'\rangle &=&
  e \langle\psi | c\bfalpha\cdot\nabla\Lambda + \dot\Lambda|\psi'\rangle.
\end{eqnarray*}
The advantage of the Dirac Hamiltonian is that the difference
$H_A- H_{A'}$ does not depend on $A$, but a similar
calculation can be carried out in the non-relativistic
case.\cite{Feuchtwang-86}
Then, we notice that
$c\bfalpha\cdot\nabla \Lambda=(i/\hbar) {[}H_D,\Lambda{]}$
for any Dirac Hamiltonian $H_D$.
Thus, if $|\psi\rangle$ and $|\psi'\rangle$ are eigenstates of 
$H_D$ with energy $E$ and $E'$, we obtain
\begin{eqnarray}
\langle \psi| H_A-H_{A'}|\psi'\rangle =
e\langle \psi|\dot\Lambda|\psi'\rangle
+ i e\frac{E-E'}{\hbar}\langle \psi|\Lambda|\psi'\rangle.
\label{HAdiff}
\end{eqnarray}

If we consider the absorption cross-section 
of a photon
of energy $\hbar\omega$, then energy  conservation
implies that $E'=E+\hbar\omega$. Thus, if 
$\Lambda$ satisfies $\dot\Lambda=-i\omega\Lambda$,
then 
$\langle \phi| H_A-H_{A'}|\phi'\rangle =0$.~\cite{Feuchtwang-86,Yang-88}
In other words, by restricting the gauge transformations
to those satisfying $\dot\Lambda=-i\omega\Lambda$,
the absorption cross-section, calculated up to first
order in perturbation theory, is gauge invariant.
However, in the resonant scattering cross-section, energy conservation
does not apply to the transition involving intermediate
states, and the cross-section is not 
gauge invariant even for those 
gauges.\cite{Zeyher-76,Yang-88,Stokes-13}

Equation~(\ref{HAdiff}) shows that the matrix elements
are also gauge invariant for a time-independent gauge transformation
and energy conserving processes (i.e. $E'=E$).
However, the gauge invariance principle is not supposed
to restrict to gauges satisfying specific constraints
such as $\dot\Lambda=-i\omega\Lambda$ or $\dot\Lambda=0$.

This rapid overview shows that, in the published semi-classical
approaches where the photon is represented by an
external potential, the transition probabilities
are not gauge-invariant  and no proposed solution
has reached general acceptance. 
Therefore, we turn now to a framework
where both electrons and photons are
quantized: quantum electrodynamics (QED).

\subsection{Quantum electrodynamics}
In QED the incident light is no
longer described by an external electromagnetic
field but by a photon, i.e. a state in a bosonic Fock space. 
Therefore, a scattering experiment is now described by 
the transition
from an initial state involving both the electronic system
in its ground state and the incident photon,
to a final state involving both the electronic system
in its (possibly) excited state and the scattered photon.
Thus, the energy of the initial and final states is
the same and, in the Schr\"odinger picture,
the gauge transformation is expressed
in terms of time-independent operators 
instead of a time-dependent function $\Lambda$.\cite{Nakanishi}
Equation~\eqref{HAdiff} suggests that transition probabilities, 
which are now described through the so-called S-matrix, could be
gauge invariant.

This is indeed the case, although a review of the literature on 
the gauge invariance of QED 
might look ambiguous because the kind of gauge transformation
considered in different works can vary.
In standard textbooks, 
``the S-matrix is gauge invariant by construction''\cite{Peskin}
because only the so-called $\xi$-term is modified.
In the most general gauge transformation, 
the space of states change
from one gauge to the other.\cite{Steinmann}
For example, in the Coulomb gauge, only the
transverse degrees of freedom are quantized and 
the photon states form a Hilbert space
built by acting on the vacuum with
creation operators of left and right polarized
photons, while in the Lorenz gauge four degrees
of freedom are quantized and the states 
(built by acting on the vacuum with creation
operators of the left, right, longitudinal and scalar 
photons) can have a negative norm.  In the Lorenz gauge, the Lorenz condition
cannot be satisfied as an operator equation,\cite{Strocchi-13}
it becomes a subsidiary condition used
to determine a subspace of physical states
with positive norm.

In other words, the state spaces of
the Coulomb and Lorenz gauges have a quite
different nature and the relation between
them is delicate. Haller managed to
show that the usual gauges are equivalent 
by devising a common framework containing
all of them.\cite{Haller-94}
Note also that the gauge-invariance can only be expected
for the renormalized 
S-matrix.\cite{Bialynicki-70,Kallosh-73,Manoukian-88}

The gauge invariance under a general \emph{infinitesimal}
gauge transformation is well established
within the Becchi-Rouet-Stora-Tyutin (BRST) approach:
matrix elements of gauge-invariant operators between physical
states are independent
of the choice of the gauge-fixing functional 
if and only if the physical states $|\alpha\rangle$
satisfy $Q|\alpha\rangle=0$, where
$Q$ is the BRST 
charge.\cite{WeinbergQFTII,Hollands-08}
The case of finite BRST
transformations is in progress.\cite{Batalin-14,Moshin-15}

To summarize the discussion, the
gauge invariance of the renormalized S-matrix is established for
infinitesimal gauge transformations and for 
a reasonably large classe of gauges.\cite{Haller-79,Matsuda-80,%
Voronov-82,Manoukian-88,Haller-94,Lenz-94,Kashiwa,Grigore-01,%
Das-13-2} In other words, it is 
proved at the physicist level of rigour.

The most studied gauges are the Lorenz and Coulomb 
gauges. Renormalization is perfectly established for the Lorenz gauge, but 
in most practical calculations the subsidiary condition
(Gauss' law) is not enforced.\cite{Haller-68}
Although it was proved that the S-matrix elements are
often the same with and without the subsidiary 
condition,\cite{Haller-70,Haller-75,Haller-79,%
Aharonov-79,Kazes-82} this fails when the Hamiltonian
is suddenly changed,\cite{Haller-69} as in the sudden
creation of a core hole in photoemission
or x-ray absorption~\cite{Ohtaka-90,Klevak-14}.
In that case, Gauss' law has to be imposed
in the Lorentz gauge and the Coulomb gauge
result is recovered.\cite{Haller-69}

We choose to use quantum electrodynamics in 
the Coulomb gauge because 
it is the most accurate gauge for low-energy
many-body calculations.\cite{Cohen-QED,Grant-07}

\section{Relativistic matrix elements}
Since we now have a gauge-invariant framework,
we can calculate the relativistic matrix elements
that will be used in
 x-ray scattering
and absorption cross-sections.

\subsection{The Hamiltonian}
The quantum field Hamiltonian describing the interaction
of light with matter in the Coulomb gauge 
is:\cite{Haller-79,GreinerFQ,Cohen-QED,Bialynicki-84}

\begin{eqnarray*}
H &=& H_e+H_\gamma + H_{e\gamma},
\end{eqnarray*}
where
\begin{eqnarray*}
H_e &=& \int d\bfr \psi^\dagger(\bfr) \big(
  c \bfalpha \cdot(-i\hbar\nabla -e\bfa)+ \beta mc^2 + e \phi
  \big) \psi(\bfr) 
\\&& + \int d\bfr d\bfr'
   \frac{\rho(\bfr)\rho(\bfr')}{8\pi \epsilon_0 |\bfr-\bfr'|},
\end{eqnarray*}
where $\phi$ is a time-independent scalar external potential
(for instance the nuclear potential),
$\bfa$ is a time-independent vector potential
(describing an external magnetic field) and 
$\psi$ are fermion field operators.
Normal ordering is implicit in $H_e$.
It is the QED form of the 
Dirac Hamiltonian in the Coulomb gauge.
The many-body version of this Hamiltonian is
\begin{eqnarray*}
H_N &=& 
\sum_{n=1}^N
  c \bfalpha_n \cdot(-i\hbar\nabla_n -e\bfa(\bfr_n))+ \beta_n mc^2 + e 
  \phi(\bfr_n)
\\&& + \sum_{m\not=n} \frac{e^2}{8\pi \epsilon_0 }
   \frac{1}{|\bfr_m-\bfr_n|},
\end{eqnarray*}
where $\bfalpha_n$ and $\beta_n$ act on the
$n$th Dirac electron. 
It can be given a well-defined mathematical
meaning if the electronic system is described with respect
to the Dirac sea,\cite{Esteban-08}
although the physical validity of the Dirac sea
is sometimes disputed.\cite{Kutzelnigg-12}

The photon Hamiltonian is
\begin{eqnarray*}
H_\gamma =  
\frac{\epsilon_0}{2}\int d\bfr
|\bfE^\perp|^2+ c^2 |\bfB|^2=
\sum_{\bfk,l} \hbar\omega_{\bfk,l} a^\dagger_{\bfk,l}  
  a_{\bfk,l},
\end{eqnarray*}
where $l$ stands for the polarization of a mode 
(there are two independent directions for a given wavevector $\bfk$)
and 
\begin{eqnarray*}
H_{e\gamma} &=&-ec \int d\bfr \psi^\dagger(\bfr) \bfalpha\cdot\bfA(\bfr)
  \psi(\bfr),
\end{eqnarray*}
describes the photon-matter interaction
in the Coulomb gauge.
According to Bialynicki-Birula, the Hamiltonian $H$ also
describes the dynamics of gauge-invariant states
in any gauge.\cite{Bialynicki-84} 
The many-body version of this interaction Hamiltonian is
\begin{eqnarray*}
H_{I} &=&-ec \sum_{n=1}^N \bfalpha_n\cdot\bfA(\bfr_n).
\end{eqnarray*}

\subsection{S-matrix elements}
Since we saw that the S-matrix is gauge invariant,
we calculate its matrix-elements.
We recall that 
\begin{eqnarray}
S &=& \lim_{\epsilon\to 0} 
  T(e^{-\frac{i}{\hbar} \int_{-\infty}^\infty H_\epsilon(t)dt}),
\label{Smatrix}
\end{eqnarray}
where $H_\epsilon(t)=e^{-\epsilon|t|} e^{iH_0t} H_{e\gamma}
e^{-iH_0 t}$. The adiabatic switching 
factor $e^{-\epsilon|t|}$ enables us
to describe physical processes as matrix elements
of $S$ between eigenstates of 
$H_0=H_e+H_\gamma$. The limit $\epsilon\to0$ can be shown
to exist up to technical assumptions.\cite{BPS}
Note that $H_0$ is not quadratic because
of the Coulomb interaction term in $H_e$.
The eigenstates of $H_e$ are
correlated multi-electronic wavefunctions.
As a consequence, we are not in the 
textbook framework, the time-dependence
of $H_\epsilon(t)$ 
cannot be calculated explicitly and the Feynman
diagram technique is no longer available 
to describe electrons.
We can bypass this problem with the
so-called ``non-covariant'' approach,\cite{Heitler} using
matrix elements of $H_\epsilon(t)$ between
eigenstates of $H_0$.
Then, cross-sections are expressed in terms
of the S-matrix and T-matrix elements
related by:
\begin{eqnarray*}
\langle m|S|n\rangle &=& \delta_{mn}
- 2i\pi\delta(e_m-e_n) \langle m|T|n\rangle.
\end{eqnarray*}
Up to second order,
\begin{eqnarray}
\langle m|T|n\rangle = 
\langle m|H_{e\gamma}|n\rangle + 
\sum_p \frac{\langle m| H_{e\gamma}|p\rangle \langle
p| H_{e\gamma}|n\rangle}
{e_p-e_n+i\gamma},
\label{Smatrixelements}
\end{eqnarray}
where $|m\rangle$, $|p\rangle$  and $|n\rangle$ are eigenstates
of $H_0$ with energy
$e_m$, $e_p$  and $e_n$, respectively.
The term $i\gamma$ was added as a heuristic way
to avoid divergence at resonance (i.e. when
the states $|n\rangle$ and $|p\rangle$ are 
degenerate). More sophisticated methods exist
to deal with such degeneracies~\cite{BPS-PRL}
but they would bring us too far. 
From the physical point of view, $\gamma$ 
describes the life-time of the state
$|p\rangle$, which can decay by 
radiative or non-radiative relaxation.
The sign of the damping term $\gamma$  has been the object
of some controversy.\cite{Veklenko-87,Hassing,%
Milonni-08,Mukamel2007}

Let us stress again that, since $H_e$ is not quadratic,
we essentially work in the Schr\"odinger picture,
where the operators are independent of time, instead
of the standard interaction picture which is used in most
textbooks.  Both approaches are 
equivalent.\cite{Epstein-55} A modern version of
the Schr\"odinger picture of QFT is given
by Hatfield.\cite{Hatfield}

Our purpose is now to calculate the matrix elements
$\langle m|H_{e\gamma}|n\rangle$,
where $H_{e\gamma}$ is independent of time.
The second quantized expression for the
photon field in the Schr\"odinger picture 
is:\cite{Strange} 
\begin{eqnarray*}
\bfA(\bfr)=\sum_{\bfk,l} \sqrt{\frac{\hbar}{2\epsilon_0V\omega_{\bfk}}} 
\left(\bfepsilon_{\bfk,l} a_{\bfk,l} \e^{i\bfk\cdot\bfr}+\bfepsilon^\star_{\bfk,l}a^\dagger_{\bfk,l} 
\e^{-i\bfk\cdot\bfr}\right).
\end{eqnarray*}
Note that we do not assume the polarization 
vectors $\bfepsilon_{\bfk,l}$ to be real.

We denote $|n\rangle=a^\dagger_{\bfk,l}|0\rangle |\Psi_n \rangle$ 
an eigenstate of 
$H_0$ where one photon is present in mode $\bfk,l$ and 
the electrons are in state $|\Psi_n\rangle$ with energy $E_n$.
The energy of $|n\rangle$ is $e_n=\hbar \omega_{\bfk,l}+E_n$.
The interaction Hamiltonian
$H_{e\gamma} $ is linear in $\bfA$ which is linear
 in photon creation and annihilation operators so that 
 only one-photon transitions are possible.
The state $|n\rangle$ can make transitions towards
$|a\rangle=|0\rangle |\Psi_m \rangle$ by absorption
and 
$|e\rangle=a^\dagger_{\bfk,l}a^\dagger_{\bfk',l'}|0\rangle| \Psi_m \rangle$ by emission.
From now on, we denote $\omega=\omega_{\bfk,l}$,
$\bfepsilon=\bfepsilon_{\bfk,l}$, $\omega'=\omega_{\bfk',l'}$ and 
$\bfepsilon'=\bfepsilon_{\bfk',l'}$.
The corresponding matrix elements are:
\begin{eqnarray*}
\langle  a\vert H_{e\gamma} \vert  n\rangle = -ec  
\sqrt{\frac{\hbar}{2\epsilon_0V\omega}}
\bfepsilon\cdot 
\langle  \Psi_m \vert   
\int\psi^\dagger \bfalpha \psi \e^{ i\bfk\cdot\bfr}\vert  
\Psi_n \rangle,
\end{eqnarray*}
and
\begin{eqnarray*}
\langle  e\vert H_{e\gamma} \vert  n\rangle = -ec  
\sqrt{\frac{\hbar}{2\epsilon_0V\omega'}} 
\bfepsilon'^\star\cdot 
\langle  \Psi_m \vert   \int\psi^\dagger \bfalpha \psi
\e^{-i\bfk'\cdot\bfr}\vert  
\Psi_n \rangle,
\end{eqnarray*}
where 
\begin{eqnarray*}
\int\psi^\dagger \bfalpha \psi\e^{\pm i\bfk\cdot\bfr} &=&
\int \psi^\dagger(\bfr) \bfalpha \psi(\bfr) \e^{\pm i\bfk\cdot\bfr}
d\bfr.
\end{eqnarray*}

\subsection{Electric dipole and multipole transitions}
To carry out a multipole expansion of the previous
matrix elements, we shall continue working with quantum fields
instead of the usual many-body expressions.
In that framework,
the expressions are simpler because there is no
electron index and we can use the following
well-known trick.\cite{Baye-12,Strocchi-13}

Let $F=\int \psi^\dagger(\bfr) f(\bfr) \psi(\bfr) d\bfr$,
where $f$ is some function of $\bfr$.
To calculate the commutator of $F$ with some
Hamiltonian $H_0$, we go to the interaction
picture and define $F_I(t)=e^{i H_0 t/\hbar}
F e^{-i H_0 t/\hbar}$. 
Then, the time-derivative $\dot{F}_I$ of $F_I$ is given by
$-i\hbar \dot{F}_I(t)={[}H_0,F_I(t){]}$.
Now, we notice that $F$ is related to the
density operator $\rho(\bfr)=\psi^\dagger(\bfr)\psi(\bfr)$ by
$F=\int \rho(\bfr) f(\bfr) d\bfr$.
Thus, 
$-i\hbar \dot{F}_I(t) = -i\hbar\int \dot{\rho}(\bfr,t) f(\bfr) d\bfr
= {[}H_0,F_I(t){]}$.
If $H_0$ conserves the electric charge, 
the continuity equation $e\dot{\rho}(\bfr)=-\nabla\cdot \bfj$
holds, where $\bfj$ is the electric current operator.
By taking $t=0$ to recover the operators in the
Schr\"odinger picture, we obtain
\begin{eqnarray}
{[}H_0,F{]} &=&  \frac{i\hbar}{e} \int \nabla\cdot\bfj(\bfr)
  f(\bfr) d\bfr  
= -\frac{i\hbar}{e} \int \bfj(\bfr) \cdot\nabla
  f(\bfr) d\bfr \nonumber 
\\&=& -i\hbar c \int \psi^\dagger(\bfr) \bfalpha \psi(\bfr)
  \cdot \nabla f(\bfr) d\bfr.
\label{H0F}
\end{eqnarray}
To find the electric dipole transition term
we apply Eq.~(\ref{H0F}) with
$f(\bfr)=\bfepsilon\cdot\bfr$ and $H_0=H_e$:
\begin{eqnarray*}
{[}H_e,\int \psi^\dagger(\bfr) \bfepsilon\cdot \bfr\psi(\bfr)d\bfr{]}
=
 -i\hbar c\int
  \psi^\dagger(\bfr) \bfalpha\psi(\bfr) \cdot\bfepsilon d\bfr,
\end{eqnarray*}
and we obtain in the dipole approximation
$\e^{i\bfk\cdot\bfr} \simeq 1$ 
\begin{eqnarray*}
\langle  a\vert H_{e\gamma} \vert  n\rangle = \frac{e(E_m-E_n)}{i\hbar}  
\sqrt{\frac{\hbar}{2\epsilon_0V\omega}}
\langle  \Psi_m \vert   \int\psi^\dagger \bfepsilon\cdot\bfr \psi\vert  
\Psi_n \rangle.
\end{eqnarray*}

To deal with electric quadrupole and magnetic dipole
transitions, we expand  to the first order:
$\e^{i\bfk\cdot\bfr} \simeq 1 + i \bfk\cdot\bfr$.
We  apply Eq.~(\ref{H0F}) with
$f(\bfr)=\bfepsilon\cdot\bfr \bfk\cdot\bfr$ and $H_0=H_e$:
\begin{eqnarray*}
{[}H_e,\psi^\dagger\bfepsilon\cdot\bfr\bfk\cdot\bfr\psi{]}
= -i\hbar c \psi^\dagger \bfalpha \psi \cdot 
  (\bfepsilon \bfk\cdot\bfr + \bfk \bfepsilon\cdot\bfr),
\end{eqnarray*}
where we removed the integral sign for notational convenience.
Thus,
\begin{eqnarray*}
\psi^\dagger \bfepsilon\cdot\bfalpha \bfk\cdot\bfr\psi
&=& \frac{i}{\hbar c} {[}H_e,
\psi^\dagger\bfepsilon\cdot\bfr\bfk\cdot\bfr\psi{]}
- \psi^\dagger\bfepsilon\cdot\bfr \bfk\cdot\bfalpha\psi.
\end{eqnarray*}
If we add $\psi^\dagger \bfepsilon\cdot\bfalpha \bfk\cdot\bfr\psi$
to both terms we obtain
\begin{eqnarray*}
2\psi^\dagger \bfepsilon\cdot\bfalpha \bfk\cdot\bfr\psi
&=& \frac{i}{\hbar c} {[}H_e,
\psi^\dagger\bfepsilon\cdot\bfr\bfk\cdot\bfr\psi{]}
\\&&
- \psi^\dagger(\bfepsilon\times\bfk)\cdot(\bfr \times\bfalpha)\psi.
\end{eqnarray*}
Finally, up to electric quadrupole transitions
\begin{eqnarray}
\langle  a\vert H_{e\gamma} \vert  n\rangle = \frac{e\Delta E}{i\hbar}  
\sqrt{\frac{\hbar }{2\epsilon_0V\omega}}
\langle  \Psi_m \vert   \int\psi^\dagger T
\psi\vert  
\Psi_n \rangle,
\label{ahn}
\end{eqnarray}
where $\Delta E= E_m-E_n$ and
\begin{eqnarray}
T =
\bfepsilon\cdot\bfr 
+ \frac{i}{2}
\bfepsilon\cdot\bfr \bfk\cdot\bfr 
-
\frac{\hbar c}{2\Delta E} (\bfepsilon\times\bfk)\cdot(\bfr
\times\bfalpha).
\label{defT}
\end{eqnarray}
The first term of $T$ is the usual electric-dipole
operator, the second one is the electric-quadrupole
operator and the third one will turn out to be the
magnetic-dipole operator (see section~\ref{semirelmul}).
Similarly,
\begin{eqnarray}
\langle  e\vert H_{e\gamma} \vert  n\rangle = \frac{e\Delta E}{i\hbar}  
\sqrt{\frac{\hbar }{2\epsilon_0V\omega'}}
\langle  \Psi_m \vert   \int\psi^\dagger T'
\psi\vert  
\Psi_n \rangle,
\label{ehn}
\end{eqnarray}
where 
\begin{eqnarray*}
T' =
\bfepsilon'^\star\cdot\bfr 
- \frac{i}{2}
\bfepsilon'^\star\cdot\bfr \bfk'\cdot\bfr 
+
\frac{\hbar c}{2\Delta E} (\bfepsilon'^\star\times\bfk')\cdot(\bfr
\times\bfalpha).
\end{eqnarray*}

\section{Semi-relativistic representation}
\label{sec:semi-relativistic}
In the previous sections, we have shown that
gauge invariance is ensured by describing
the interaction of light and matter
with quantum electrodynamics, where photons
are quantized and electrons
are described by four-component Dirac spinor quantum fields.

However, in most solid-state calculations, we
do not use Dirac spinors but two-component
(Pauli) wavefunctions. Moreover, semi-relativistic
expressions are often physically clearer.
Therefore, we need 
to link the two representations by using
a generalization of the Foldy-Wouthuysen transformation.

In this section, we first describe the Foldy-Wouthuysen transformation
and its new many-body extension. Then, we use this framework
to calculate the relativistic corrections to the dipole
and quadrupole transitions. The calculations are considerably
simpler than the usual approach, where the relativistic corrections
are derived from a semi-relativistic Hamiltonian.

\subsection{The Foldy-Wouthuysen transformation}
The idea of the Foldy-Wouthuysen transformation is the
following. If $H_D$ is a time-independent 
relativistic Hamiltonian, it has the form
\begin{eqnarray*}
H_D &=& H^0 + \left(\begin{array}{cc} 
   H_{11}  & H_{12}  \\ H_{21}  & H_{22} \end{array}\right),
\end{eqnarray*}
where $H^0=mc^2\beta$ and each $H_{ij}$ is a 2x2 matrix.
We write $H_D$ as the sum of even and odd parts
$H_D=H^0+\calE + \calO$, where
\begin{eqnarray*}
\calE = \left(\begin{array}{cc} 
   H_{11}  & 0  \\ 0  & H_{22} \end{array}\right),\quad
\calO = \left(\begin{array}{cc} 
   0  & H_{12}  \\ H_{21}  & 0 \end{array}\right),
\end{eqnarray*}
satisfy $\beta\calE\beta=\calE$ and
$\beta\calO\beta=-\calO$. Note that $H^0$ is also even.
If $|\psi_D\rangle$ is a solution of the Dirac equation
$H_D|\psi_D\rangle=E|\psi_D\rangle$, where
$H_D$ is the Dirac Hamiltonian, then
the upper two components of $|\psi_D\rangle$ are called the
large components and the lower two the small components.
The Dirac equation 
couples the large and small components
of $|\psi_D\rangle$ through the odd terms of $H_D$.
Foldy and Wouthuysen~\cite{Foldy-50} 
looked for a unitary operator $U$ that decouples
the large and small components of 
$|\psi\rangle=U|\psi_D\rangle$.
In other words, $H=U H_D U^\dagger$ 
has only even components:
$H=\beta H \beta$.
The method proposed by Foldy and Wouthuysen consists
in successive transformations of the form $U=e^{iS}$.
~\cite{Foldy-50,GreinerRQM}.

This
transformation does not satisfy
Eriksen's condition $U=\beta U^\dagger \beta$
discussed in the Appendix. This is because 
the product
$U=e^{iS^{(2)}} e^{iS^{(1)}}$  does not
satisfy this equation even if
$e^{iS^{(1)}}$ and $e^{iS^{(2)}}$ do.
Silenko recently derived the correction that must
be applied to go from Foldy-Wouthuysen to
Eriksen transformations,\cite{Silenko-16} and he showed that
the correction is at an order beyond the one 
we consider in this paper.

\subsection{Many-body Foldy-Wouthuysen transformation}
To generalize the Foldy-Wouthuysen approach to
the many-body Dirac Hamiltonian we face the
following problem.
The generalization of $H^0$ is imposed by the
many-body Dirac Hamiltonian:
\begin{eqnarray*}
H^0_N &=& \sum_{n=1}^N \beta_n mc^2,
\end{eqnarray*}
where $\beta_n$ is the matrix $\beta$ acting
on the $n$th electron
(i.e. $\beta_n=1^{\otimes (n-1)}\otimes\beta\otimes 
1^{\otimes (N-n)}$). This 
definition is valid because $H^0_N$ commutes
with the projector $P_N$ onto the space of antisymmetric
$N$-body states.

We show in the Appendix that a Foldy-Wouthuysen
transformation can be defined whenever we have a
self-adjoint operator $\eta$ (with $\eta^2=1$)
to define parity.
In the one-body case, $\beta^2=1$ and $\eta=\beta$ defines
parity. But in the many-body case
the operator $\sum_{n=1}^N \beta_n$ suggested by
$H^0_N$ cannot be used for that purpose
because its square is not proportional to the identity
(it contains products $\beta_n\beta_m$). It turns out
that $\eta=\beta_1\otimes\dots\otimes\beta_n$
is the natural many-body generalization of $\beta$.
Indeed, $\eta^\dagger=\eta$ and $\eta^2=1$.
Moreover, $\eta$ commutes with $P_N$, which allows
us to work with tensor products instead of 
antisymmetric tensor products.

In the literature, the Foldy-Wouthuysen transformation was
studied for two-body 
Hamiltonians,\cite{Chraplyvy-53,Chraplyvy-53-2,Eriksen-58},
but the results were rather complicated 
and not easy to extend to the many-body case.

The even and odd parts of 
$H_N$ are then $H^0_N+\calE$ and $\calO$,
respectively:
\begin{eqnarray*}
\calE&=&  e \sum_{n=1}^N \phi_0(\bfr_n)
 + e  \sum_{m\not=n}  V(\bfr_m-\bfr_n),\\
\calO &=& 
\sum_{n=1}^N
  c \bfalpha_n \cdot\bfpi_n=
\sum_{n=1}^N \calO_n,
\end{eqnarray*}
where $V(\bfr)=\frac{e}{8\pi \epsilon_0 |\bfr|}$
is the Coulomb potential and $\bfpi_n=-i\hbar\nabla_n -e\bfa_0(\bfr_n)$.


At first order in $c^{-1}$,  the Foldy-Wouthuysen
operator is $U=e^{iS^{(1)}}$  where
\begin{eqnarray*}
S^{(1)} &=& - \frac{i}{2mc^2} \sum_n \beta_n \calO_n.
\end{eqnarray*}
Indeed, it can be checked that
$i{[}S^{(1)},H^0_N{]}=-\calO$ removes the odd term
of $H_D$.  At this order
$U=U_1\otimes\dots\otimes U_N$ is a tensor
power of one-body Foldy-Wouthuysen operators,
as proposed by Moshinksy and Nikitin.\cite{Moshinsky-04}

However, this tensor-power form does not hold at higher orders.
Indeed, we show now that at the next order, the many-body Foldy-Wouthuysen
Hamiltonian is the sum of
one-body and two-body contributions.
The usual formal Foldy-Wouthuysen
transformation $U=e^{iS^{(1)}}e^{iS^{(2)}}$  can be carried out almost unchanged
and we find, with $m$ as expansion parameter, at order $m^{-2}$:
\begin{eqnarray*}
H_\mathrm{FW} &=& H_N^0 + \calE + \frac{1}{2mc^2} \sum_{n=1}^N 
\beta_n\calO_n^2 
\\&&
-
 \frac{1}{8m^2c^4} \sum_{n=1}^N 
{[}\calO_n,{[}\calO_n,e\varphi_n+eV{]} {]}
\\&&
+ 
 \frac{1}{8m^2c^4} \sum_{p\not= n}
\beta_p\beta_n {[}\calO_p,{[}\calO_n,V{]}{]}.
\end{eqnarray*}
This Hamiltonian obeys $\eta H_\mathrm{FW} \eta =H_\mathrm{FW} $ which makes it a Foldy-Wouthuysen Hamiltonian. 

It rewrites
\begin{equation}
H_\mathrm{FW}= \sum_{n=1}^N  H_{\mathrm{FW}}^{n}+H_{\mathrm{FW}}^{MB}.
\end{equation}
where each $H_{\mathrm{FW}}^{n}$ is the 
usual one-body Foldy-Wouthuysen Hamiltonian:
\begin{multline*}
H_\mathrm{FW}^{n} = \beta_n m c^2+ e  \phi_0(\bfr_n)
 + \sum_{p\not=n} e V(\bfr_n-\bfr_p) \\ 
 + \frac{1}{2m} \beta_n  
 \bfpi_n^2-e \hbar \bfSigma_n \cdot \bfb_0(\bfr_n) 
 -  \frac{\hbar^2 e}{8m^2c^2} \nabla \cdot\bfE_{n} \\
  +\frac{\hbar e}{8m^2c^2} \bfSigma_n \cdot 
   ( \bfpi_n \times \bfE_{n} - \bfE_{n} \times \bfpi_n )
\end{multline*}
where 
\begin{equation*}
\bfb_0(\bfr_n)= \nabla \times \bfa_0(\bfr_n)
\end{equation*}
and
\begin{equation*}
\bfE_{n}=- \nabla \phi_n(\bfr_p)-\sum_{p\neq n} \nabla V(\bfr_n-\bfr_p).
\end{equation*}
The mass-velocity term $\frac{\beta_n}{8m^3c^2} (\bfp_n\cdot \bfp_n)^2$ 
would be obtained by expanding to higher order.

The new two-body term $H_{\mathrm{FW}}^{MB}$ arises because 
$V(\bfr_m-\bfr_n)=V(\bfr_{mn})$
is a two body operator:
\begin{multline*}
H_{\mathrm{FW}}^{n,p} = \frac{\hbar e}{8m^2c^2} \sum_{p\neq n}^N  \Big(\hbar \Delta V(\bfr_{np}) \\
 - \bfSigma_n \cdot (\bfpi_n  \times \nabla V(\bfr_{np})  - \nabla V(\bfr_{np}) \times \bfpi_n )\\
 +2  \hbar  \beta_n \beta_p
(\bfalpha_n \cdot \nabla_n)(\bfalpha_p \cdot \nabla_p) V(\bfr_{np}) \Big).
\end{multline*}

By using:\cite{Weiglhofer}
\begin{eqnarray*}
\partial_j\partial_k V(\bfr) &=& 
\frac{e^2}{8\pi\epsilon_0} 
\Big(-\delta_{jk} \frac{4\pi}{3}\delta(\bfr)
-\delta_{jk} \frac{1}{r^3}
+ \frac{3 r^j r^k}{r^5}\Big),
\end{eqnarray*}
the derivatives in the last term can be rewritten
\begin{align*}
(\bfalpha_n \cdot \nabla_n)(\bfalpha_p \cdot \nabla_p)
V(\bfr_{np}) &=\sum_{jk} \bfalpha_n^j \bfalpha_p^k  \partial_j\partial_k V(\bfr_{np})\\
=\frac{e^2}{8\pi\epsilon_0} \Big(-\frac{4\pi}{3}\bfalpha_n\cdot\bfalpha_p \delta(\bfr_{np})&
- \frac{\bfalpha_n\cdot\bfalpha_p}{|\bfr_{np}|^3}
\\
&+ 3\frac{\bfalpha_n\cdot\bfr_{np}
   \bfalpha_p\cdot\bfr_{np}}{|\bfr_{np}|^5}\Big).
\end{align*}
This expression looks superficially like
some contributions to the Breit 
interaction as presented by Bethe
and Salpeter.\cite{Bethe-Salpeter}
However, they are different since the
Breit interaction is due to the exchange
of a photon and not to a semi-relativistic
effect.
Note that the last two terms are singular.
It is known that the expansion of the
Foldy-Wouthuysen transformation as a power
serie in $1/c^2$ becomes
more and more singular because of the presence
of the Coulomb potential.\cite{Morrison}
At order $m^{-2}$, the transformation writes
\begin{multline*}
U=1 + \frac{1}{2mc^2}\sum_n \beta_n \calO_n - \frac{1}{8m^2c^4}\Big( \sum_n \beta_n \calO_n \Big)^2 \\
+\frac{1}{4m^2c^4} \sum_n \beta_n \Big[\sum_m \beta_m \calO_m ,\calE\Big]
\end{multline*}
and it obeys $U=\eta U^\dagger \eta$.
We also checked that $U^2$ is odd in $H_D$
after paying attention to the discontinuity
at zero discussed in the Appendix.
Thus, the positive (negative) energy eigenstate of $H_D$ are
transformed into even (odd) states by the action of $U$.

\subsection{Semi-relativistic dipole transitions}
Matrix elements such as 
$D=\langle  \Phi \vert   \int\psi^\dagger \bfepsilon\cdot\bfr \psi\vert  
\Psi \rangle$ 
are now evaluated by expressing the positive energy Dirac wavefunctions
$|\Phi\rangle$ and $|\Psi\rangle$
in terms of the Foldy-Wouthuysen ones $|\phi\rangle$ and
$|\psi\rangle$:
$|\Phi\rangle=U^\dagger|\phi\rangle$ and
$|\Psi\rangle=U^\dagger|\psi\rangle$.
Since $U$ is written as a many-body operator,
we translate the quantum field expression for $D$
into the many-body formula
$D=\langle \Phi|\bfepsilon\cdot\bfR |\Psi\rangle$,
where $\bfR=\sum_{n=1}^N \bfr_n$.\cite{Fetter,Gross}
We calculate
$D=\langle \phi|U\bfepsilon\cdot\bfR U^\dagger|\psi\rangle$,
where $U=e^{iS}$ by using the Baker-Campbell-Hausdorff formula
\begin{eqnarray*}
e^{iS} T e^{-iS} &=& T +i [S,T] + \sum_{n=2}^\infty i^n
   \frac{L^n(T)}{n!},
\end{eqnarray*}
where $L (T)=[S,T]$ and $L^n(T)=L(L^{n-1}(T))$.
If $U=U_1\otimes\dots\otimes U_N$, where
$U_i=e^{iS_i}$, we
can calculate the action of $U$ on each variable
independently. Removing temporarily the
constant $-i/2mc^2$, we take the
one-body operator $S=\beta\calO$ and compute
\begin{eqnarray*}
L(\hat\epsilon\cdot\bfr) &=& 
c{[} \beta\bfalpha\cdot(\bfp-e\bfa_0),\hat\epsilon\cdot\bfr{]} =
c\sum_{ij} \beta\alpha^i \epsilon^j
 {[}p_i,r_j{]} 
\\&=&
-i\hbar c\sum_{ij} \beta\alpha^i \epsilon^j \delta_{ij}
=  -i\hbar c \beta\bfalpha\cdot\hat\epsilon,
\end{eqnarray*}
and
\begin{eqnarray*}
L^2(\hat\epsilon\cdot\bfr) &=& 
-i\hbar c^2 {[} \beta\bfalpha\cdot(\bfp-e\bfa_0),
\beta\bfalpha\cdot\hat\epsilon{]} 
\\&=& -i\hbar c^2 \sum_{ij} 
   (p_i-ea_{0i})\epsilon_j {[}\beta\alpha^i,\beta\alpha^j{]}
\\&=&
 i\hbar c^2 \sum_{ij} 
   (p_i-ea_{0i})\epsilon_j {[}\alpha^i,\alpha^j{]},
\end{eqnarray*}
where we used $\beta\alpha_i=-\alpha_i\beta$ and $\beta^2=1$.
We compute
\begin{eqnarray*}
{[}\alpha^i,\alpha^j{]} &=& 
 2 i \sum_k \epsilon_{ijk} 
 \left(\begin{array}{cc} 
   \sigma^k & 0 \\ 
  0 & \sigma^k \end{array}\right)
=2 i \sum_k \epsilon_{ijk} \Sigma^k,
\end{eqnarray*}
which defines $\Sigma^k$ the components of $\bfSigma$.
Therefore,
\begin{eqnarray*}
L^2(\hat\epsilon\cdot\bfr) &=& 
 -2\hbar c^2 
   (\bfp-e\bfa_0)\cdot(\hat\epsilon\times\bfSigma).
\end{eqnarray*}
So that, for each particle, and up to $O(m^{-2})$,
\begin{eqnarray*}
U_n \bfepsilon\cdot\bfr_n U_n^\dagger &=&
   \bfepsilon\cdot\bfr_n -i \frac{\hbar}{2mc}
   \beta_n\bfalpha_n\cdot\bfepsilon
\\&&
 - \frac{\hbar}{4m^2c^2} \bfpi_n
   \cdot(\bfepsilon\times\bfSigma_n).
\end{eqnarray*}
The many-body version is obtained by summing the
right-hand side over $n$.

In the matrix elements
$D=\langle \phi|U\bfepsilon\cdot\bfR U^\dagger|\psi\rangle$,
recall that $|\psi\rangle=\eta|\psi\rangle$ and
$|\phi\rangle=\eta|\phi\rangle$
because $|\Psi\rangle$ and $|\Phi\rangle$ are positive
energy states, as shown in the Appendix.
Therefore,
$\langle \phi|U\bfepsilon\cdot\bfR U^\dagger|\psi\rangle=
\langle \phi|\eta U\bfepsilon\cdot\bfR U^\dagger\eta|\psi\rangle$
and all the terms that are odd
in $U\bfepsilon\cdot\bfR U^\dagger$ are eliminated
by the matrix elements.
This eliminates the term
proportional to $\beta_n\bfalpha_n$ and we are left with
\begin{eqnarray*}
D = \sum_{n=1}^N 
\langle\phi|\bfepsilon\cdot\bfr_n
 - \frac{\hbar}{4m^2c^2} 
\bfpi_n
   \cdot(\bfepsilon\times\bfSigma_n)|\psi\rangle.
\end{eqnarray*}

\subsection{Semi-relativistic multipole transitions}
\label{semirelmul}
From Eq.~(\ref{defT}), we write the multipole
transitions
\begin{eqnarray*}
M &=&\frac{i}{2}M_1 - \frac{\hbar c}{2\Delta E}M_2,
\end{eqnarray*}
where
\begin{eqnarray*}
M_1 &=&\sum_n
\langle \phi|U\bfepsilon\cdot\bfr_n \bfk\cdot\bfr_nU^\dagger|\psi\rangle,
\\
M_2 &=&\sum_n
\langle\phi|U(\bfepsilon\times\bfk)\cdot(\bfr_n
\times\bfalpha_n)U^\dagger|\psi\rangle,
\end{eqnarray*}
correspond to the electric quadrupole and
magnetic dipole transitions, respectively.
Since multipole transitions are smaller than dipole
ones, it is enough to use the first two
terms of the Baker-Campbell-Hausdorf formula.

The term
${[}S_n,\bfepsilon\cdot\bfr_n
\bfk_n\cdot\bfr_n{]}$ 
is odd and disappears  in the matrix element.
Thus, at the order we consider,
\begin{eqnarray*}
M_1 &=&\sum_n
\langle \phi|\bfepsilon\cdot\bfr_n \bfk\cdot\bfr_n|\psi\rangle.
\end{eqnarray*}
Let $T_2=(\bfepsilon\times\bfk)\cdot(\bfr
\times\bfalpha)$. We write
\begin{eqnarray*}
{[}\beta\calO,T_2 {]} &=&
c {[}\beta\bfalpha \cdot \bfp,T_2{]}-e c{[}\beta\bfalpha \cdot \bfa_0,T_2{]}\\
&=&c\beta ( \lbrace\bfalpha \cdot \bfp,T_2 \rbrace-
e  \lbrace\bfalpha \cdot \bfa_0,T_2\rbrace).
\end{eqnarray*}
The anticommutators are 
\begin{align*}
\lbrace\bfalpha \cdot \bfp,T_2 \rbrace &=  
\sum_{ijkl} \epsilon_{jkl} (\bfepsilon \times \bfk)_j (\alpha_i \alpha_l p_i  r_k  + r_k p_i  \alpha_l \alpha_i ) \\
&=2 (\bfepsilon \times \bfk) \cdot (\hbar  \bfSigma + \bfL),
\end{align*}
and 
\begin{align*}
 \lbrace\bfalpha \cdot \bfa_0,T_2\rbrace  &= \sum_{ijkl}\epsilon_{ikl}
(\bfepsilon \times \bfk)_i {a}_j r_k \lbrace \alpha_j , \alpha_l \rbrace\\
 &=2 (\bfepsilon \times \bfk) \cdot (\bfr \times \bfa_0).
\end{align*}
Note that $\hbar\bfSigma=g\bfS$ with $g=2$
(because the spin operator is
$\bfS=\hbar\bfSigma/2$). Thus, we recover the
fact that the Dirac equation gives a gyromagnetic factor
$g=2$ to the electron. Moreover,
$\bfL+\hbar\bfSigma=\bfL+2\bfS$ is the total
magnetic moment of the electron.

Finally, since $\bfr_n\times\bfalpha_n$ is odd,
\begin{eqnarray*}
M_2 &=&\sum_n \frac{\beta_n}{mc}
\langle\phi|(\bfepsilon\times\bfk)\cdot
(\hbar  \bfSigma_n + \bfLambda_n)|\psi\rangle,
\end{eqnarray*}
where $\bfLambda_n=\bfL_n-e\bfr_n\times\bfa_0(\bfr_n)$
is the moment of the mechanical momentum 
as defined in Ref.~\onlinecite{CDL}. 
The term $M_2$ describes magnetic-dipole transitions.
The multipole transitions are
\begin{eqnarray*}
M &=& \sum_n \langle\phi| \frac{i}{2}
\bfepsilon\cdot\bfr_n \bfk\cdot\bfr_n
\\&&
-\frac{\hbar\beta_n}{2m\Delta E}(\bfepsilon\times\bfk)\cdot
(\hbar  \bfSigma_n + \bfLambda_n)|\psi\rangle.
\end{eqnarray*}

\section{Absorption cross-section}
The absorption cross section is calculated by assuming that initially
the system of electrons is in state $|I\rangle$ that can be transformed into
 Foldy-Wouthuysen eigenstate $|i\rangle$,
   with energy $E_i$, and that a photon $\bfk,\bfepsilon$ is present. In the final state
there is no photon and the system is in state $|F\rangle$ ($|f\rangle$ after transformation).

The transition probability per unit time from state
$m$ to state $n$ is related to 
the T-matrix elements by:\cite{Walecka}
\begin{equation}
w= \frac{2}{\hbar} \delta_{mn}
  \Im \langle m |T|m\rangle
+ \frac{2\pi}{\hbar} \delta(e_n-e_m) |\langle n|T|m\rangle|^2.
\label{wS}
\end{equation}
and must be divided by $c/V$ (rate at which the photon crosses a unit of surface)
 to obtain the cross section.
Since we consider real transitions
(i.e. $m\not= n$), only the second term is present.

From \eqref{ahn} and using the result of transformation derived in the previous section:
\begin{eqnarray*}
\sigma &=& 4\pi^2\alpha_0\hbar\omega
\sum_f |\langle f|T_{\mathrm{FW}}|i\rangle|^2
\delta(E_f-E_i-\hbar\omega),
\end{eqnarray*}
where $T_{\mathrm{FW}}$ is:
\begin{eqnarray*}
T_{\mathrm{FW}} &=& \sum_n \bfepsilon\cdot\bfr_n 
+
\frac{i}{2}
\bfepsilon\cdot\bfr_n \bfk\cdot\bfr_n
 - \frac{\hbar}{4m^2c^2} 
\bfpi_n \cdot(\bfepsilon\times\bfSigma_n)
\\&&
-\frac{\beta_n}{2m\omega}(\bfepsilon\times\bfk)\cdot
(\hbar  \bfSigma_n + \bfLambda_n),
\end{eqnarray*}
with $\alpha_0$ the fine structure constant
and $\Delta E= E_f-E_i=\hbar\omega$.

It corresponds to the usual formula for the 
cross section \cite{Brouder-96-XMCD} with two more terms: 
the third one and the last one.

The third term 
was already found by Christos Gougoussis in 
his PhD thesis,\cite{GougoussisPhD}
but his final result was not in agreement with ours
because of his use of the commutation relation, as described
in section~\ref{Commut-sect}.
We rewrite it by using 
$\bfpi=(m/i\hbar) [\bfr,H_0^\mathrm{FW}]+O(c^{-2})$, where
$H_0^\mathrm{FW}$ is the Foldy-Wouthuysen Hamiltonian, to get:
\begin{align*}
- \frac{\hbar}{4m^2c^2}
\langle f|&\bfpi\cdot(\bfepsilon\times\bfSigma)|i\rangle\\&=
 \frac{i}{4mc^2} (E_i-E_f)
\langle f|\bfr\cdot(\bfepsilon\times\bfSigma)|i\rangle
\\&=
\frac{i\hbar\omega}{4mc^2}
\langle f|(\bfepsilon\times\bfr)\cdot\bfSigma|i\rangle.
\end{align*}
We call \emph{spin-position operator} 
the operator $(\bfepsilon\times\bfr)\cdot\bfSigma$.
Its evaluation at the K-edge of materials will be
presented in a companion paper.\cite{Bouldi-XMCD}

The amplitude of the last term depends on the 
choice of the space origin in the Coulomb gauge  for $\bfa_0$.
It does not make the cross section gauge dependent
because the states are changed accordingly when choosing
the origin of the gauge.
If the origin of the gauge is chosen at the atom position,
fields larger than $10^6$ T
are required for this term to be significant.
Such fields are way beyond laboratory accessible
values.

\section{Scattering cross-section}
The scattering cross section is calculated by assuming that initially
the system of electrons is in state $|I\rangle$ with a photon 
$\bfk_i,\bfepsilon_i$ and that in the final state the system 
is in state $|F\rangle$ with a scattered photon $\bfk_f,\bfepsilon_f$. 
We do not consider the special case when $\bfk_i,\bfepsilon_i=\bfk_f,\bfepsilon_f$.

Eqs.~\eqref{ahn},~\eqref{ehn} and \eqref{wS}  yield:
\begin{multline*}
w = \frac{2\pi}{\hbar}\sum_F \delta(E_f+\hbar \omega_f-E_i-\hbar \omega_i)\Big| \sum_L \frac{ e^2c^2\hbar}{2\epsilon_0V} \frac{1}{\sqrt{\omega_i\omega_f}}\\
\frac{\langle
F|e_{-\bfk_f}\psi^\dagger \bfalpha\cdot \bfepsilon_f^\star\psi|L\rangle
\langle L|e_{\bfk_i}\psi^\dagger \bfalpha\cdot\bfepsilon_i\psi
|I\rangle}{E_i-E_l+\hbar \omega_i+i\gamma} \\
+\frac{\langle F|e_{\bfk_i}\psi^\dagger \bfalpha\cdot\bfepsilon_i\psi
|L\rangle\langle L|e_{-\bfk_f}\psi^\dagger \bfalpha\cdot
\bfepsilon_f^\star\psi|I\rangle}{E_i-E_l-\hbar \omega_f} 
  \Big|^2,
\end{multline*}
where $\gamma>0$ and
\begin{eqnarray*}
e_{\bfk}\psi^\dagger \bfalpha\cdot \bfepsilon\psi
&=& \sum_{j=1}^3 \int e^{i\bfk\cdot\bfr}
  \psi^\dagger(\bfr) \alpha^j \psi(\bfr) \epsilon^j d\bfr.
\end{eqnarray*}
The scattering cross-section is related to $w$
by:\cite{AlsNielsen}
\begin{eqnarray*}
\frac{d^2 \sigma}{d\Omega d \omega_f}
&=&
\frac{V^2}{(2\pi)^3} \omega_f^2 \frac{1}{\hbar c^4} w.
\end{eqnarray*}
Since the electric charge is related to the 
classical electron radius $r_e$ by
$e^2=4\pi \epsilon_0 r_e m c^2$, we obtain
the relativistic Kramers-Heisenberg scattering cross-section:
\begin{eqnarray*}
\frac{d^2 \sigma}{d\Omega d \omega_f}  &=& 
  (r_e mc^2)^2 \frac{\omega_f}{\omega_i}
  \sum_F \delta(E_f+\hbar \omega_f-E_i-\hbar \omega_i)
    \nonumber\\ &&\Big|\sum_L
\frac{\langle
F|e_{-\bfk_f}\psi^\dagger \bfalpha\cdot \bfepsilon_f^\star\psi|L\rangle
\langle L|e_{\bfk_i}\psi^\dagger \bfalpha\cdot\bfepsilon_i\psi
|I\rangle}{E_i-E_l+\hbar \omega_i+i\gamma} \nonumber\\
&+&\frac{\langle F|e_{\bfk_i}\psi^\dagger \bfalpha\cdot\bfepsilon_i\psi
|L\rangle\langle L|e_{-\bfk_f}\psi^\dagger \bfalpha\cdot
\bfepsilon_f^\star\psi|I\rangle}{E_i-E_l-\hbar \omega_f} 
  \Big|^2.
\end{eqnarray*}
In this expression, the 
sum over  $|L\rangle$ involves a complete
set of states, with positive and negative energies.
Since $E_i$ is usually the 
positive energy of the ground state
including the electron rest energy,
we have $E_i=mc^2+E'_i>0$, where $E'_i$
is the usual (negative) ground state energy.
If $|L\rangle$ is a positive energy state,
we have $E_l=mc^2 +E'_l$ with $E'_l > E_i'$
and the first term is resonant at
$\hbar\omega_i=E'_l-E'_i$.
If $|L\rangle$ is a negative energy state,
then $E_l=-mc^2-E'_l$
and $E_i-E_l-\hbar\omega_f=2 mc^2 + E'_i-E'_l 
-\hbar\omega_f$ cannot be resonant 
in standard experimental conditions.

We show that the resonant scattering term has
a semi-relativistic expansion close to, but different from, the
standard one.~\cite{Blume-85}
If we are interested in the resonant part of the
scattering cross section, then $E_l>0$ and
\begin{eqnarray*}
\frac{d^2 \sigma}{d\Omega d \omega_f}  &=& 
  (\frac{r_e m}{\hbar^2})^2 \frac{\omega_f}{\omega_i}
  \sum_f \delta(E_f+\hbar \omega_f-E_i-\hbar \omega_i)
    \nonumber\\ &&\Big|\sum_{L^>}
(E_l-E_i)(E_f-E_l)\\ 
&&\frac{\langle
f|T'^{fl}_{\mathrm{FW}}(\bfepsilon_f)|l\rangle\langle l|
T^{li}_{\mathrm{FW}}(\bfepsilon_i)|i\rangle}{E_i-E_l+\hbar \omega_i+i\gamma}
\Big|^2.
\end{eqnarray*}
 with
\begin{eqnarray*}
T^{ij}_{\mathrm{FW}}(\bfepsilon_i) &=&  \sum_n\bfepsilon_i\cdot\bfr_{n} 
+
\frac{i}{2}
\bfepsilon_i\cdot\bfr_{n} \bfk_i\cdot\bfr_{n}
\\&&
 - \frac{\hbar}{4m^2c^2} 
\bfpi_{n} \cdot(\bfepsilon_i\times\bfSigma_{n})
\\&&
-\frac{\hbar\beta_n}{2m\Delta E^{ij}}(\bfepsilon_i\times\bfk_i)\cdot
(\hbar  \bfSigma_{n} + \bfLambda_{n}),
\end{eqnarray*} 
 and 
 \begin{eqnarray*}
T'^{ij}_{\mathrm{FW}}(\bfepsilon_f) &=&  \sum_n\bfepsilon_f^\star\cdot\bfr_{n} 
-
\frac{i}{2}
\bfepsilon_f^\star\cdot\bfr_{n} \bfk_f\cdot\bfr_{n}
\\&&
 - \frac{\hbar}{4m^2c^2} 
\bfpi_{n} \cdot(\bfepsilon_f^\star\times\bfSigma_{n})
\\&&
+\frac{\hbar\beta_n}{2m\Delta E^{ij}}(\bfepsilon_f^\star\times\bfk_f)\cdot
(\hbar  \bfSigma_{n} + \bfLambda_{n}),
\end{eqnarray*} 
where $\Delta E^{ij}=E_i-E_j$.

As in the absorption case, the spin-position term in the transition
operator is not present in the usual formula.\cite{Blume-85} 

\section{Other methods}
In this section, we compare our semi-relativistic
transition matrix elements with the ones obtained
by using time-dependent perturbation theory where the time-evolution
is described by several time-dependent semi-relativistic
Hamiltonians: the one proposed by 
Blume, the  ``gauge-invariant'' Foldy-Wouthuysen
one, the textbook Foldy-Wouthuysen one 
and the effective Hamiltonian derived in
non-relativistic QED (NRQED). 
Before making this comparison, we first 
explain why using a time-dependent semi-relativistic
Hamiltonian in a perturbation calculation
can lead to incorrect results.

\subsection{Foldy-Wouthuysen subtelties}
\label{FWsub-sect}
In this section, we assume that the exact
time-dependent Foldy-Wouthuysen operator 
$U$ is known. Thus, the following difficulties are
not related to the use of an approximation, but to
the interplay of the Foldy-Wouthuysen method 
with perturbation theory.

The  first subtelty was
noticed by Nieto:\cite{Nieto-77,Goldman-77}
If $|\Psi\rangle$ is a solution of the time-dependent 
Dirac equation $(i\hbar\partial_t -H)|\Psi\rangle=0$, 
then the Foldy-Wouthuysen transformation turns it into
$|\psi\rangle=U|\Psi\rangle$, where
$U$ is a unitary time-dependent operator.
The time-dependent Dirac equation for 
$|\Psi\rangle$ implies that 
$|\psi\rangle$ is a solution of the time-dependent
Schr\"odinger equation 
$(i\hbar\partial_t - H')|\psi\rangle=0$, where 
$H'=UHU^{-1}+ i \hbar (\partial_t U)U^{-1}$
is the time-dependent
Foldy-Wouthuysen Hamiltonian.
In the following, 
an uppercase Greek letter ($|\Phi\rangle$
or $|\Psi\rangle$) refers to a solution of the 
Dirac equation and the corresponding lowercase letter 
($|\phi\rangle$
or $|\psi\rangle$) to its Foldy-Wouthuysen
transformation.

As a consequence, a matrix element
$\langle \Phi|H|\Psi\rangle$ is not equal to
$\langle \phi |H'|\psi\rangle$, but to
$\langle \phi |H'-i \hbar (\partial_t U)U^{-1}|\psi\rangle$.
In other words, $H'$ has to be used to calculate
the states $|\phi\rangle$ and $|\psi\rangle$ but not
to calculate the matrix elements of the Hamiltonian.

The second subtelty was observed by Yang.\cite{Yang-82-3} and concerns the most straightforward
way to use the Foldy-Wouthuysen Hamiltonian $H'(t)$,
where the time dependence is now explicit, to compute transition
probabilities. 
This Hamiltonian is split into a time-independent part
$H'_0$ and a time-dependent one $H'_1(t)$, so that
$H'(t)=H'_0+H'_1(t)$. The scalar product $\langle \phi'_n|\psi(t)\rangle$, where
$|\phi'_n\rangle$ is an eigenstate of $H'_0$, cannot be equal
to the relativistic scalar product 
$\langle \Phi_n|\Psi(t)\rangle$.  Indeed
$|\psi(t)\rangle=U(H_0+H_1(t))|\Psi(t)\rangle$
but $|\phi'_n\rangle\not=U(H_0+H_1(t))|\Phi_n\rangle$
because $|\phi'_n\rangle$ and $|\Phi_n\rangle$
are independent of time whereas
$U(H_0+H_1(t))$ depends on time.
Since only the QED relativistic matrix elements where
found to be gauge invariant, 
$\langle \phi'_n|\psi(t)\rangle$ is generally not physically
meaningful.

The two problems combine if first-order perturbation theory
is naively applied with Foldy-Wouthuysen eigenstates and Hamiltonian. 
The Foldy-Wouthuysen interaction Hamiltonian
$H'_1(t)=H'(t)-H'_0\neq U(H_0) (H(t)-H_0) U^{\dagger}(H_0)$. 
As a consequence, 
$\langle \phi_n' |H_1'(t)|\phi_g' \rangle$
is not equal to $\langle \Phi_n |H_1(t)|\Phi_g \rangle$.

To illustrate the variety of results that can be obtained
by using first-order perturbation theory with 
semi-relativistic Hamiltonians, we now examine four Hamiltonians
used in practice.
To help comparing these Hamiltonians, 
we express them in a common one-particle framework.

\subsection{The Blume Hamiltonian}

Blume discussed the interaction of light with
magnetic matter by starting from the 
Hamiltonian:\cite{Blume-83,Blume-85}
\begin{eqnarray}
 \HB = \frac{\bfpi^2}{2m}
 + eV- \frac{e \hbar}{2m}\mathbf{\bfsigma}\cdot\bfB 
 -\frac{e\hbar}{4m^2c^2}\bfsigma \cdot 
   (\bfE\times \bfpi ),
   \label{HBlume}
\end{eqnarray}
where $\bfpi=\bfp -e \bfA$.
This Hamiltonian is the sum of four terms:
(i) the kinetic energy of the electron,
(ii) an external potential,
(iii) the Zeeman interaction between the electron and a
 magnetic field and (iv) the spin-orbit interaction
(because, for a spherical $V$ and a static $\bfA$,
$\bfsigma \cdot (\bfE \times \bfp)=\frac{-1}{r}\frac{dV}{dr}
\bfsigma\cdot(\bfr\times\bfp)=\frac{-1}{r}\frac{dV}{dr}
\bfsigma\cdot\bfL$).

There are several differences between our notation and
Blume's: he considers a many-body Hamiltonian 
(involving sums over electrons)
and writes $\sum_{ij} V(r_{ij})$ for our
$eV$, he adds the Hamiltonian $H_\gamma$ 
of the free photons, he uses $\bfA/c$, $\nabla\times\bfA/c$ 
and $\bfs$ where we use $\bfA$, $\bfB$
and $\bfsigma/2$, 
finally, his Zeeman term is wrong by a factor of 2
in his first two papers on the subject,\cite{Blume-83,Blume-85}
but this was corrected in the third one.\cite{Blume-94}
In this third paper, Blume also replaces
$\bfE$ by $-\dot{\bfA}$. This is not compatible
with his quantized description of the photon field.
Indeed, the time-derivative $\dot\bfA$ is present
in the Lagrangian but, after the Legendre transformation
leading to the Hamiltonian, $\dot\bfA$ is replaced
by its canonical momentum $-\bfE$.
Note that Blume does not sketch any derivation
of his Hamiltonian.

\subsection{Foldy-Wouthuysen Hamiltonian}
We consider now the so-called ``gauge-invariant''
Foldy-Wouthuysen Hamiltonian for positive-energy
states up to order $1/(mc)^2$:\cite{Foldy-52}
\begin{eqnarray*}
\HFW = \HB
 + m c^2  -\frac{e\hbar^2}{8m^2c^2}  \nabla\cdot\bfE 
-\frac{ie\hbar^2}{8m^2c^2}\bfsigma \cdot (\nabla\times\bfE).
\end{eqnarray*}

The difference between the Foldy-Wouthuysen and
the Blume Hamiltonians consists of three terms:
the rest energy $mc^2$ of positive-energy
eigenstates, the Darwin term proportional to
$\nabla\cdot\bfE$ and a last term,
proportional to $\bfsigma \cdot (\nabla\times\bfE)$
and called the \emph{curl-term}, that we discuss presently.
A basic difference between $\HB$
and $\HFW$ must first be stressed: 
the former is a QED expression
where the quantum fields $\bfA$, $\bfB$ and $\bfE$
are independent of time because they are
written in the Schr\"odinger representation,
while the latter was derived under the assumption that
$\bfA$ and $V$ are external time-dependent potentials.
In particular, the curl-term disappears if
the external field $\bfA$ is independent of time.\cite{Strange}
In the semi-classical treatment of light-matter interaction,
the photons are represented by an external time-dependent
potential and this term is present.

These Hamiltonians can be written
$H(\bfA,\Phi)$, where the total vector potential $\bfA$
and scalar potential $\Phi$ are a sum
$\bfA=\bfa_0+\bfa$, $\Phi=\phi_0+\phi$, 
of static external potentials
$\bfa_0$ and $\phi_0$ (representing the
static internal and external fields) perturbed by 
dynamical potentials $\bfa$ and $\phi$
representing the incident electromagnetic wave.
We write the interaction Hamiltonian
as $H_I=H(\bfA,\Phi)-H(\bfa_0,\phi_0)$.
The two Hamiltonians $\HB$ and $\HFW$ lead to
two different interactions:
$\HB_I=h_1+h_2+h_3+h_4+h_5+h_6$
and $\HFW_I=\HB_I+h_7$, where
\begin{eqnarray*}
h_1&=&\frac{e^2}{2m} \bfa^2,\\
h_2&=&-\frac{e}{m} \bfa\cdot \bfpi_0,\\
h_3&=&-\frac{e\hbar}{2m} \bfsigma\cdot(\nabla \times \bfa),\\
h_4&=&\frac{e^2\hbar}{4 m^2c^2} \bfsigma\cdot(\bfe \times \bfa), \\
h_5&=&-\frac{e\hbar}{4 m^2c^2} \bfsigma\cdot(\bfe \times \bfpi_0 ), \\
h_6&=&\frac{e^2\hbar}{4 m^2c^2} \bfsigma\cdot(\bfe_0 \times \bfa),\\
h_7&=&-\frac{ie\hbar^2}{8m^2c^2}\bfsigma \cdot(\nabla\times \bfe) ,
\end{eqnarray*}
with $\bfpi_0=\bfp-e\bfa_0$. The \emph{curl-term} in $\HFW $
is the origin of the presence of $h_7$ in $\HFW_I$. It
originates from the term
$i \hbar (\partial_t U)U^{-1}$ in the time-dependent
Foldy-Wouthuysen Hamiltonian.
The Darwin term gives no contribution to the interaction
because $\nabla\cdot\bfe$ is zero for the electromagnetic
wave.
The terms $h_5$ and $h_6$ were omitted by Blume,
who considered them to be small.\cite{Blume-83} 
We shall see that $h_5$ is the source of a spin-position
term which is not negligible in x-ray magnetic circular dichroism
(XMCD) spectra.\cite{Bouldi-XMCD}

\subsection{Textbook Foldy-Wouthuysen Hamiltonian}

Standard textbooks often derive a
Foldy-Wouthuysen Hamiltonian $H^{\mathrm{TFW}}$
which is the same as $H^{\mathrm{FW}}$, except for the fact that
$\bfpi$ is replaced by $\bfp$ in the
spin-orbit term.\cite{Itzykson,Bjorken}
A mass-velocity term $-(\bfp\cdot\bfp)^2/8m^3c^2$ is 
often added \cite{Itzykson} but its contribution 
to the radiation-matter interaction is zero.
The difference with the Foldy-Wouthuysen
Hamiltonians is a term in $\bfsigma \cdot \bfE \times \bfA$.
This results in the absence of $h_4$ and $h_6$ in the 
perturbation Hamiltonian, which changes the 
transition probabilities.

\subsection{NRQED}
To deal with QED calculations involving bound states,
Caswell and Lepage proposed an alternative
approach to relativistic effects, called
non-relativistic QED (NRQED),  which 
turned out to be highly successful.\cite{Caswell-86}
They wrote the most general gauge-invariant 
non-relativistic Lagrangian terms and fitted the coefficients of
these terms to known QED processes.\cite{Paz-15} 

The corresponding NRQED Hamiltonian
is the same as $\HFW$ up to order $c^{-2}$, but its interpretation
is different.\cite{Paz-15} Indeed, NRQED is a quantum field theory, 
and the fields are independent of time in the 
Schr\"odinger representation.
However, the curl-term is present in time-independent NRQED although it
is generated by a time-dependence in $H^{\mathrm{FW}}$.
In particular, the curl-term must not be removed from the
Hamiltonian to calculate matrix elements of the 
Hamiltonian operator, in contrast to the  example of
section~\ref{FWsub-sect}.

Besides these four different Hamiltonians,
we consider an additional source
of discrepancies between authors: the commutators.

\subsection{Commutators}
\label{Commut-sect}
To derive the multipole expansion of the matrix element of $H_I$,
it is useful to replace $\bfpi$ by a commutator with 
$H_0=H(\bfa_0,\phi_0)$.
The derivations that start from Blume's interaction 
Hamiltonian usually use the relation:\cite{Takahashi-15,Joly-12}
\begin{equation}
\bfp=\frac{mi}{\hbar}[H_0,\bfr].
\label{commutH0}
\end{equation}
However, if one considers the static Hamiltonian given by 
Blume \eqref{HBlume}, its commutator with $\bfr$ is:
\begin{equation*}
[\HB_0,\bfr]=-\frac{i\hbar}{m}\bfpi_0+
\frac{e\hbar}{4m^2c^2}(i\hbar)(\bfsigma\times \bfe_0),
\end{equation*}
which is different from Eq.~\eqref{commutH0} because $\bfp$ is 
replaced by $\bfpi_0=\bfp-e\bfa_0$ and because of 
the term proportional to $c^{-2}$.
The commutator of $\bfr$ with $H_0^{\mathrm{TFW}}$ and $\HFW_0$
are the same.
In  $\HFW_I$ and  $\HB_I$, when $\bfpi_0$ in $h_2$ is rewritten 
as a function of the commutator, the extra relativistic term 
leads to the cancellation of $h_6$, which is
important in XMCD.
On the other hand, it leads to a contribution
 $\frac{e^2\hbar}{4 m^2c^2} \bfsigma\cdot[\nabla v_0 \times \bfa] $
in $H_I^{\mathrm{TFW}}$.
 
If the mass-velocity term $-(\bfp\cdot\bfp)^2$ is present in $H_0$, 
the additional contribution to the commutator, 
$\frac{i\hbar(\bfp\cdot\bfp)\bfp}{2m^3c^2}$ is small compared 
to $\frac{i\hbar}{m}\bfp$ if the order of magnitude of the
 kinetic energy of the core state satisfies $E_k<<mc^2$.

For all the Hamiltonians presented here, using the relation 
$[\bfp,v_0]=i\hbar \nabla v_0$, the electric
field in matter writes at zeroth order in $c^{-2}$ as a function
of the commutator of $\bfpi_0$ with $H_0$:
\begin{equation*}
\bfe_0=-\nabla v_0=\frac{-i}{e\hbar}[\bfpi_0,H_0].
\end{equation*}
In the case of absorption, the commutator transforms into 
a factor $\Delta E=-\hbar \omega$ in the cross section so that 
$h_5$ and $h_6$ lead to the same contribution to the matrix element:
\begin{equation*}
\frac{-ie\hbar \omega}{4m^2c^2} \bfsigma \cdot ( \bfa \times \bfpi_0),
\end{equation*} 
which corresponds to the \emph{spin-position} interaction.
Explicit calculations showed that this contribution can
appear  two times, one time or cancel completely,
according to which Hamiltonian and which commutator
was used.
Starting from $\HFW$, the same absorption
cross section as in our new approach can be derived. However,
in the case of scattering, even with $\HFW$, there is a factor
$\Delta E/\hbar \omega$ which is not correct.
The same kind of discrepancy was already observed 
in the literature.\cite{Zeyher-76,Yang-88}

\section{Conclusion}
This paper was written because of the 
gauge-dependence of transition probabilities
in the semi-classical approach and because
we observed, after other authors,\cite{Takahashi-15,DiMatteo-16}
that different semi-relativistic
Hamiltonians lead to different 
cross-sections.

Our solution makes essential use of
quantum electrodynamics as the correct
gauge-invariant framework to discuss the
interaction of light with matter.
It is well-known that the semi-classical 
and QED absorption cross-sections are identical in the
Coulomb gauge.\cite{Loudon-00} 
This is compatible with our discussion because, to go from the
Coulomb gauge to another gauge, the semi-classical
approach only involves the operator 
$M_\Lambda$, while QED involves a redefinition
of the space of states, including 
in an essential way non-physical polarizations
and even ghost states in the BRST approach.
This redefinition is able to maintain gauge
invariance where the semi-classical $M_\Lambda$
fails to do so.

In the present paper, the stationary states of
the electronic system was taken to be eigenstates
of $H_e$. The interaction Hamiltonian $H_{e\gamma}$
can modify these states through various QED effects, for example 
the Breit interaction discussed by
Bethe and Salpeter.\cite{Bethe-Salpeter}
We expect these contributions to
be small in x-ray spectroscopy. 

The explicit calculation of the spin-position
contribution at the K-edge of Fe, Co and Ni
will presented in a forthcoming
publication.\cite{Bouldi-XMCD}

It was known since Heisenberg in 1928,\cite{Brown-02}
that the Thomson cross-section
which is due to the $A^2$ term in the non-relativistic
approach, can be derived from the relativistic
framework by using a sum over
negative-energy states.\cite{Sakurai,Strange}
We intend to provide a more accurate discussion
of the contribution of negative-energy states 
to the scattering cross-section by using
our many-body Foldy-Wouthuysen approach.


\begin{acknowledgments}
We are very grateful to Uwe Gerstmann, Matteo Calandra
and Nora Jenny Vollmers for encouraging us to
work on the problem of the relativistic effects
in x-ray absorption spectroscopy.
We thank Alexander Silenko for this help
concerning the Foldy-Wouthuysen transformations.
Discussions with Am\'{e}lie Juhin,  Sergio Di Matteo, 
Yves Joly and Philippe Sainctavit
are gratefully acknowledged.
We are very grateful to Maria Esteban for her guidance
through the mathematical literature on the many-body
Dirac equation.

This work was supported by French state funds managed by the ANR within
the Investissements d'Avenir programme under Reference No.
ANR-11-IDEX-0004-02, and more specifically within the framework of the
Cluster of Excellence MATISSE led by Sorbonne Universit{\'e}s.
\end{acknowledgments}

\appendix*
\section{General Foldy-Wouthuysen transformation}
To derive a many-body Foldy-Wouthuysen transformation,
we first notice that, in the one-body case,
$\beta$ endows the space of spinors with
the structure of a Krein space, where
$\beta$ is then called a \emph{fundamental symmetry}.\cite{Baum-81}
For quite a different purpose,\cite{Brouder-15-NCG}
we investigated the tensor product of such 
spaces and showed that the fundamentaly symmetry
of the $N$th tensor power is essentially 
$\eta=\beta^{\otimes N}$.
The abstract Krein-space framework leads us 
naturally to the following theorem:

\emph{Assume that $H_D$ and $\eta$ are self-adjoint operators
and $\eta^2=1$. Then, there is a unitary
operator $U$ such that $U=\eta U^\dagger \eta$
and $\eta U H_D U^\dagger\eta=U H_D U^\dagger$.
Moreover, if $|\psi_D\rangle$ is an eigenstate of 
$H_D$ with positive (resp. negative) eigenvalue,
then $|\psi\rangle=U|\psi_D\rangle$ satisfies
$|\psi\rangle=\eta|\psi\rangle$ 
(resp. $|\psi\rangle=-\eta|\psi\rangle$).}

The condition $U=\eta U^\dagger \eta$ does not
appear in Foldy and Wouthuysen works.
It was added by
Eriksen.~\cite{Eriksen-58,Eriksen-60,Silenko-16}
It means that $U$ is self-adjoint for the Krein-space
structure.

Let us start with general considerations involving
a self-adjoint operator $\eta$ such that $\eta^2=1$.
It can be used to define projectors
$B_\pm =(1\pm\eta)/2$. It is clear that
$B_++B_-=1$,
$B_\pm^2=B_\pm$, $B_\pm^\dagger=B_\pm$ and
$B_+ B_- = B_- B_+=0$.
A vector $|\psi\rangle$ is said to be
even (odd) if $\eta|\psi\rangle=|\psi\rangle$
($\eta|\psi\rangle=-|\psi\rangle$).
Then, any vector $|\psi\rangle$ can be written as
the sum of its even part $B_+|\psi\rangle$
and its odd part $B_-|\psi\rangle$. 
An operator $H$ is said to be even (odd) if
it transforms an even state into an even
(odd) state and an odd state into an odd
(even) state.
An operator $H$ is even (odd) if and only if
$\eta H \eta=H$ ($\eta H \eta=-H$).
Thus, the theorem states that
$U H_D U$ is an even operator.
Any operator $H$ can be written as the
sum of its even part $B_+ H B_+ + B_- H B_-$
and its odd part $B_+ H B_- + B_- H B_+$.

Our proof of the theorem is essentially
a generalized
and rigorous version of Eriksen's proof.\cite{Eriksen-60}
We use the fact that $H_D$ is self-adjoint
to define $\lambda=\sign H_D$ by functional calculus.
The operator $\lambda$ is called the \emph{flat band Hamiltonian}
in topological insulator theory.\cite{Prodan-16}
In physical terms, let $|\psi_D\rangle$ be an eigenstate of $H_D$ for
the energy $E$, then 
$\lambda |\psi_D\rangle=|\psi_D\rangle$ if $E\ge 0$ and
$\lambda |\psi_D\rangle=-|\psi_D\rangle$ if $E < 0$.
Since $\eta$ and $\lambda$ are self-adjoint
and $\eta^2=\lambda^2=1$, they are bounded and
$\eta\lambda$ is unitary:
$\eta\lambda(\eta\lambda)^\dagger=\eta\lambda\lambda\eta=
\eta^2=1$ and
$(\eta\lambda)^\dagger\eta\lambda=1$.
By the spectral theorem for unitary 
operators,\cite{Pourahmadi}
there is a unique family of orthogonal projections $P_t$ such that
\begin{eqnarray*}
\eta\lambda &=& \int_{-\pi}^\pi e^{it} P_t dt.
\end{eqnarray*}
In the finite dimensional case we could write this~\cite{Lang-Lin-Alg}
\begin{eqnarray*}
\eta\lambda &=& \sum_n e^{i t_n} |\phi_n\rangle\langle\phi_n|.
\end{eqnarray*}
Thus,
\begin{eqnarray*}
\lambda\eta &=& (\eta\lambda)^\dagger = \int_{-\pi}^\pi e^{-it} P_t dt
= \int_{-\pi}^\pi e^{it} P_{-t} dt,
\end{eqnarray*}
and, by unicity of $P_t$, 
$\eta\lambda=\eta(\eta\lambda)^\dagger\eta$ implies
$P_t=\eta P_{-t}\eta$.
We can now define a unitary square
root $U$ of $\eta\lambda$
by functional calculus:~\cite{Wallen-69,Joye-14}
\begin{eqnarray*}
U &=& \sqrt{\eta\lambda}=\int_{-\pi}^\pi e^{it/2} P_t dt,
\end{eqnarray*}
which satisfies
\begin{eqnarray*}
\eta U^\dagger \eta &=& 
\int_{-\pi}^\pi e^{-it/2} \eta P_t \eta dt=
\int_{-\pi}^\pi e^{-it/2} P_{-t} dt=U.
\end{eqnarray*}

We now show that this $U$ satisfies the
intertwining relation $\eta U=U\lambda$.
Indeed, the relation $U^2=\eta\lambda$ implies
$U=U^\dagger\eta\lambda$. By multiplying
from the left with $\eta$ and using 
$\eta U^\dagger \eta=U$ we find
$\eta U=U\lambda$.
This important relation implies that 
$H=\eta H \eta$ and that 
$|\psi\rangle = U|\psi_D\rangle$ is
even 
if $|\psi_D\rangle$ is a positive
energy state and odd 
if $|\psi_D\rangle$ is a negative
energy state.

The first property is easy to show:
\begin{eqnarray*}
\eta H \eta &=& \eta U H_D U^\dagger \eta
= U\lambda H_D \lambda U^\dagger
= UH_D \lambda^2 U^\dagger
=H,
\end{eqnarray*}
because $\lambda$ commutes with $H_D$ since
it is a function of $H_D$.

To show the second property, let $\Gamma_\pm=(1\pm\lambda)/2$,
so that $\Gamma_+$ projects
onto the space of positive energy and $\Gamma_-$
of negative energy, and recall that
$B_\pm=(1\pm\eta)/2$.
For a one-body system,
$B_\pm$ projects onto the large/small components.
Then, $U\Gamma_\pm=U/2\pm U\lambda/2=U/2 \pm \eta U/2=
 B_\pm U$, which can be used to show that the 
Foldy-Wouthuysen
wavefunctions $|\psi\rangle=U|\psi_D\rangle$ 
corresponding to positive energy have only
even components.
Indeed, let $|\psi_D\rangle$ be an eigenstate of 
$H_D$ corresponding to a positive energy.
By definition of $\lambda$ we have
$\Gamma_+ |\psi_D\rangle= |\psi_D\rangle$ and
$\Gamma_- |\psi_D\rangle= 0$.
Thus, $U\Gamma_+|\psi_D\rangle=U|\psi_D\rangle=|\psi\rangle$
and $U\Gamma_+=B_+ U$ implies 
$|\psi\rangle=B_+ U |\psi_D\rangle=B_+|\psi\rangle$.
Thus $\eta|\psi\rangle=\eta
B_+|\psi\rangle=B_+|\psi\rangle=|\psi\rangle$
and $|\psi\rangle$ is even.
Similarly
$0=B_-|\psi\rangle$, so that the
odd part of $|\psi\rangle$ is zero. 

For a one-body system, 
even components 
and large components are identical. Indeed a Dirac one-body wavefunction can be written
\begin{eqnarray*}
|\psi_D\rangle &=& 
 \left(\begin{array}{c}  \phi \\ \psi \end{array}\right),
\end{eqnarray*}
If $\eta=\beta$, then the even part and the odd parts
of $|\psi_D\rangle$ are, respectively,
\begin{eqnarray*}
 \left(\begin{array}{c}  \phi \\ 0 \end{array}\right)\quad
\text{and}\quad
 \left(\begin{array}{c}  0 \\ \psi \end{array}\right).
\end{eqnarray*}
so that the small components of $|\psi\rangle$
are zero for a positive-energy state.
This is not true for many-body systems. For example, 
if we neglect antisymmetrization for notational convenience,
a two-body state can be obtained as the
tensor product of one-body wavefunctions:
\begin{eqnarray*}
|\psi_D\rangle &=& 
 \left(\begin{array}{c}  \phi_1 \\ \psi_1 \end{array}\right)
\otimes
 \left(\begin{array}{c}  \phi_2 \\ \psi_2 \end{array}\right).
\end{eqnarray*}
Then, the even part of $|\psi_D\rangle$ is 
\begin{eqnarray*}
 \left(\begin{array}{c}  \phi_1 \\ 0 \end{array}\right)
\otimes
 \left(\begin{array}{c}  \phi_2 \\ 0 \end{array}\right)
+
 \left(\begin{array}{c}  0 \\ \psi_1 \end{array}\right)
\otimes
 \left(\begin{array}{c}  0 \\ \psi_2 \end{array}\right),
\end{eqnarray*}
while its odd part is
\begin{eqnarray*}
 \left(\begin{array}{c}  \phi_1 \\ 0 \end{array}\right)
\otimes
 \left(\begin{array}{c}  0 \\ \psi_2 \end{array}\right)
+
 \left(\begin{array}{c}  0 \\ \psi_1 \end{array}\right)
\otimes
 \left(\begin{array}{c}  \phi_2 \\ 0 \end{array}\right).
\end{eqnarray*}


The characterization of $U$ as the square root of
$\eta\lambda$ is not easy to handel. 
We give now a much simpler characterization:

\emph{Let $U$ be a unitary operator continuously
defined (outside zero) in terms
of $H_D$ such that: (i) $U=\eta U^\dagger \eta$;
(ii) $\eta U H_D U^\dagger \eta=U H_D U^\dagger$;
(iii) $U^2(-H_D)=-U^2(H_D)$.
Then $U^\dagger\eta U=\pm\sign(H_D)$.}

To prove this, define $Z=U^\dagger \eta U$.
Clearly, $Z^\dagger=Z$. 
Moreover, $Z$ is defined in terms 
$H_D$ since $U$ is. 
However, for $Z$ to be a function of
$H_D$ in the sense of functional calculus,
$Z$ needs to commute with
$H_D$:\cite{Berezin-91}
if we multiply condition $(ii)$ 
from the right by $\eta U$ we find
$\eta U H_D =U H_D U^\dagger\eta U=U H_D Z$.
Hence,
\begin{eqnarray*}
Z H_D &=& U^\dagger \eta U H_D 
= U^\dagger U H_D Z=H_D Z.
\end{eqnarray*}
Thus, there is a real function $f(t)$ and a family
of orthogonal projections $P_t$ corresponding
to the eigenstates of $H_D$ such that
\begin{eqnarray*}
Z &=& \int_{-\infty}^\infty f(t) dP_t.
\end{eqnarray*}
Moreover, $Z^2=1$ because
$Z^2=U^\dagger \eta UU^\dagger \eta U=
U^\dagger \eta^2 U=
U^\dagger U=1$.
Therefore, $f^2(t)=1$ for every $t$.
Finally, observe that
$Z=\eta^2 U^\dagger \eta U=\eta U^2$, 
and condition $(iii)$ implies that $Z$ is an
odd function of $H_D$: $f(-t)=-f(t)$.
To conclude that $f(t)=\pm\sign t$, 
we need to add the condition of continuity
on $f$ outside zero. Indeed, functional calculus is valid
for measurable functions and we could
build a non-continuous odd function $f$ such
that $f^2=1$ outside the origin. In practice this does not take
place because $U$ is smoothly defined 
in terms of $H_D$, except at zero. 
No odd continuous function can satisfy
$f^2=1$ over $\mathbb{R}$. It has to be discontinuous
at zero. Since it is crucial that $f^2=1$
everywhere, we can choose
either $\sign 0=1$ or $\sign 0=-1$. 
Both solutions are valid.


\end{document}